\documentclass[]{aa}

\usepackage[varg]{txfonts}
\usepackage{graphicx,rotating,natbib}
\usepackage{amsmath,amssymb}
\bibpunct{(}{)}{;}{a}{}{,}

\begin{document}

\title{Thermal evolution and sintering of chondritic planetesimals}

\subtitle{III. Modelling the heat conductivity of porous chondrite material}

\author{ 
  Stephan Henke\inst{1}
 \and 
    Hans-Peter Gail\inst{1} 
 \and Mario Trieloff\inst{2,3}
}

\institute{
Institut f\"ur Theoretische Astrophysik, Zentrum f\"ur Astronomie, 
           Universit\"at Heidelberg, Albert-Ueberle-Str. 2,
           69120 Heidelberg, Germany 
\and
Institut f\"ur Geowissenschaften, Universit\"at Heidelberg, Im Neuenheimer
           Feld 236, 69120 Heidelberg, Germany
\and
Klaus-Tschira-Labor f\"ur Kosmochemie, Universit\"at Heidelberg, Im Neuenheimer Feld 236, 69120 Heidelberg, Germany
}

\offprints{\tt gail@uni-heidelberg.de}

\date{Received date ; accepted date}

\abstract
{
The construction of models for the internal constitution and the temporal evolution of  large planetesimals, the parent bodies of chondrites, requires an as accurate as possible information on the heat conductivity of the complex mixture of minerals and iron metal found in chondrites. The few empirical data on the heat conductivity of chondritic material are severeley disturbed by impact induced micro-cracks modifying the thermal conductivity.
}
{It is attempted to evaluate the heat conductivity of chondritic material by theoretical methods.
}
{ 
The average heat conductivity of a multi-component mineral mixture and granular medium is derived from the heat conductivities of its mixture components. Random mixtures of solids with chondritic composition and packings of spheres are generated numerically. The heat conduction equation is solved in high spatial resolution for a test cube filled with such matter. From the calculated heat flux through the cube the heat conductivity of the mixture is derived.  
}
{
For H and L chondrites, our results are in accord with empirical thermal conductivity at zero porosity. However, the porosity dependence of heat conductivity of the granular material built from chondrules and matrix is at odds with measurements for chondrites, while our calculations are consistent with data for compacted sandstone. The discrepancy is traced back to subsequent shock modification of the currently available meteoritic material by impacts on the parent body over the last 4.5 Ga. This causes a structure of void space made of fractures/cracks, which lower the thermal conductivity of the medium and acts as a barrier to heat transfer. This latter structure is different from the one probably existing in the pristine material where voids were represented by pores rather than fractures. The results obtained for the heat conductivity of the pristine material are used for calculating models for the evolution of the H chondrite parent body which are fitted to the cooling data of a
  number of H chondrites. The fit to the data is good, likewise it is with models assuming different porosity. This is an indication that more diagnostic meteorite data are needed to distinguish between porosity models.
}
{}


\maketitle


\section{%
Introduction}

Meteorites preserve in their structure and composition rather detailed information on the processes that were acting during planetary formation, in particular on processes active in their respective parent bodies. It is possible to recover this information by modelling the structure, composition, and thermal history of meteorite parent bodies and by comparing this with results of laboratory investigations of meteorites with regard to their composition and  structure, and to their thermal evolution as inferred from closure temperatures and ages of radioactive decay systems, from cooling rates. The basis of such investigations are models for the thermal evolution of their parent bodies.

The parent bodies of the ordinary chondrites are suspected to be undifferentiated planetesimals with diameters ranging between at least  hundred kilometers and at most a few hundred kilometers. There exists also the possibility that part of the parent bodies of chondrites is differentiated in their interior region and only bear an outer crust of chondritic material, see \citet{Elk11}. Irrespective of such possible structural differences the thermal evolution of such bodies is characterized by a brief initial heating period due to decay of short-lived radioactives (mainly $^{26\!}$Al), lasting for a period of a few million years, followed by a long lasting cooling period of at least 100\,Ma duration, where the heat liberated within the body is transported to its surface where it is radiated away. A realistic modelling of this raise and fall of temperature requires to use an as accurate as possible value of the heat conduction coefficient, $K$, for solving the heat conduction equation.

The present paper is the fourth in a series where we attempt to improve the model construction for the thermal evolution of planetesimals \citep{Hen11,Hen13,Gai14b}. Here we aim to improve the modelling of the heat conductivity, $K$, of the material. A determination of $K$ is complicated because of the rather complex structure of the chondrite material. First, it is a mixture of many different materials with quite different individual properties, and second it is of granular structure consisting of a two-component porous mixture of chondrules of 0.1 to 1\,mm size and a very fine-grained matrix material with particle sizes of the order of micrometers. The proportions of chondrules and matrix are strongly varying between different meteorite classes.  This granular material is found in meteorites of the petrologic classes 1 to 3 and is assumed to represent the initial structure of the material from which the parent bodies formed. 

During the course of the thermal evolution of the parent body the initially porous granular material is compacted once the temperature has raised to sufficiently high values that creep processes in the solids are activated. This results in a gradual deformation of the granules and finally in a closure of all pores. This process is witnessed by the meteorites of petrologic classes 4 to 6 (or 7) which represent different preserved stages of the compaction process at increasingly higher temperature.

In the completely compacted state the material still is a heterogeneous mixture of several mineral compounds and iron and/or iron sulphide. The next thermal evolutionary step will be the onset of melting. Meteorites where the material has once been molten, at least partially, are the achondrites or the primitive achondrites. For the ordinary chondrite classes no achondrites seem to exist that could clearly be related by compositional similarities to one of the parent bodies of the ordinary chondrite classes. Either the parent bodies of the ordinary chondrites never were heated to such an extent in their interior that their central region was molten during part of their thermal evolution, or they never suffered very strong collisions with other bodies by which material from their interior region is excavated. Generally it is assumed that the first alternative holds, but this is not really proven.
 
For studying the thermal evolution of the parent bodies of the ordinary chondrites one has to determine the heat conductivity of the granular material from its likely initial stage as a loosely packed powder or sand-like material through its various compaction stages to the final completely compacted stage of a strongly heterogeneous mixture of several components. If also melting occurs, one also needs the heat conductivity of the melt and of the material after its solidification, but this is a much simpler problem than that before melting; we do not consider bodies for which this is relevant.  

One way to determine the heat conductivity is to measure the heat conductivity coefficient for a number of chondrites and temperatures, as it was done by \citet{Yom83} and \citet{Ope10,Ope12}. Then one can use such results in model calculations. This is the way how $K$ is determined in most published model calculations of the thermal evolution of meteoritic parent bodies, including our owns. The shortcoming of this approach is, that $K$ depends on a number of parameters that change during the evolution of the bodies, and that not all stages of evolution of all the bodies are represented in the meteorite collections. 
Furthermore, a lot of cracks formed in the mineral components by shock waves, e.g. from the impact that excavated the meteorite. Such cracks reduce the heat conductivity of the material in an unpredictable way, such that measured heat conductivities of meteoritic material are not necessarily representative for the same material during the early evolution of the body. This makes it desirable to develop an approach which allows to calculate $K$ from the properties of the individual components from which the material is composed. 

Another approach is to calculate the heat conductivity from properties of the mixture components by using approximate mixing rules that have been developed for technical applications \citep[see, e.g., the review ][]{Ber95}. This has many advantages because this often combines a simple calculation method with acceptable accuracy. We will compare such methods with our more detailed calculations of the heat conductivity. 

In this paper we aim to calculate $K$ for composite materials encountered in the parent bodies of ordinary chondrites by numerically constructing samples of such composite materials and solving the equation of heat transfer for the boundary value problem of a sample held at different fixed temperatures on opposite sides of a sample. From the calculated stationary heat flux through the sample one readily determines an effective heat conductivity of the composed material. This allows to determine a value of $K$ for given composition and structure of the composite material. Though this procedure is numerically demanding, for applications it is sufficient to generate a table for the dependence of $K$ on temperature and porosity for the mixture of interest, from which $K$ can be determined by interpolation during the course of model calculations.

The heat conductivity is determined by three processes, the transport via phonons in electrically insulating minerals, by conduction electrons in the
metallic iron component of the mixture, and by radiative transfer through the
transparent minerals and the voids of the mixture. These have different dependencies on temperature and their contribution to the total heat conduction coefficient is of different order of magnitude. At low temperatures conduction by the solid material dominates and the contribution by radiative transfer is negligible. At high temperatures ($T>1\,000$\,K) the radiative contribution steeply increases and finally becomes the dominating mode of heat transfer. Both modes of heat transfer, by phonons and conduction electrons at one hand, and by radiation on the other hand, are independent of each other and can simply be superposed. 

The problem of heat conduction through minerals by radiative transfer was studied already for modelling the structure of planets. Here one can take recourse to what has been done in planetary physics \citep[e.g.][]{Hof99,Hof14} and we will not elaborate on this part of the problem. We attempt here to calculate the heat conduction by solid-state heat conduction for a material that is an intimate mixture of different mineral and iron grains and --- before compaction --- of voids. 

The voids in the granular material are filled with gas from the accretion disk and possibly with outgassed volatiles. This also contributes to the heat conductivity, but in earlier test calculations for the models in \citet{Hen11} we found this to be negligible compared to the other processes. But this may be different for carbonaceous chondrites (not considered here) during the phase of ice melting and evaporation.

The main dependencies of interest for our problem are the temperature dependence
of $K$ and its dependence on the composition and the granular structure of the
material, and how this is determined from the properties of the different
components. A possible pressure dependence is of minor importance, because
central pressures in the bodies that we have in mind do not much exceed 1\,kbar or stay well below that value. Effects resulting from the compressibility of the material remain negligible in that case. We do not consider them.

The plan of this paper is as follows: In Sect.~\ref{SectCompChon} we discuss the composition of chondrites, in Sect.~\ref{SectHeatCond} we describe the calculation of heat conductivty of composite materials and in Sect.~\ref{SectGran} that for granular materials. In Sect.~\ref{SectParBod} we apply our results to the parent body of H chondrites. The paper closes with some final remarks in Sect.~\ref{SectConclu}. The appendices contain descriptions of our calculational methods for solving the heat conduction equation, for generating random multicomponent mixtures, and to generate packing of spheres.


\begin{table*}[t]
\caption{Granular components and some properties of chondrite groups.\tablefootmark{1}}

{\small
\begin{tabular}{lcr@{}lr@{}lr@{}lr@{}lr@{}lrrrrrrr}
\hline
\hline
\noalign{\smallskip}
Chondrite group & & EH & & EL & & H & & L & & LL & & CI & CM & CR\tablefootmark{2} & CO & CK & CV & K
\\
\noalign{\smallskip}
\hline
\noalign{\smallskip}
Chondrules\tablefootmark{3}  & [vol\%] & 60-80  & & 60-80  & & 60-80 & & 60-80 & & 60-80 & &  $\ll1$& 20    & 50-60   & 48   & 15      & 45   & 25
\\[0.1cm]
Avg. diameter\tablefootmark{3}   & [mm]    & 0.2    & & 0.6    & & 0.3   & & 0.7   & & 0.9   & & -      & 0.3   & 0.7     & 0.15 & 0.7     & 1.0   & 0.6
\\[0.1cm]
Matrix\tablefootmark{3}    & [vol\%] & $<2$-15&?& $<2$-15&?& 10-15 & & 10-15 & & 10-15 & & $>99$  & 70    & 30-50   & 34   & 75       & 40   & 73
\\[0.1cm]
$f_{\rm ma}$ [Eq.~(\ref{DefFracMat})]      &       & 0.1    & & 0.1    & &  0.15   &&  0.15   &&  0.15  && 1.0 &  0.78 & 0.42   & 0.42  & 0.83    & 0.47  &  0.74 
\\[0.1cm]
Porosity\tablefootmark{6}   & [\%]      &  3-12?     & &   3-12? & &  2-12  &&  2-11   &&  5-14   &&  35  &  18-28  &   & 2-20  & 20-23  & 20-24  &       \\
$\phi_0$ (Eqs. (\ref{AppBinGran}), (\ref{AppBinGranMa}))   & [\%]      &   29   & &   29  & &  24   &&  24   &&  24   &&  36  &  30     &  19  &  19  & 32  & 21  & 30          
\\[0.1cm]
FeNi metal\tablefootmark{3,4}          & [vol\%] &       8&?&      15&?& 10    & & 5     & & 2     & & 0      & 0.1   & 5-8     & 1-5   & $<0.01$ & 0-5   & 6-9
\\[0.1cm]
Avg. diameter\tablefootmark{8}                     & mm &  0.04 &&  0.1 && 0.19 && 0.16 && 0.14 &&
\\[0.1cm]
FeS\tablefootmark{7}                                 & [vol\%] &  4-8    & &       & & 4-5    & &  4-5   & & 4-5  &  &  4    &  1-4   & 1-4   &  1-5   &  0-1    &  2-5   & 6-10 
\\[0.1cm]
CAIs\tablefootmark{3}            & [vol\%] &   0.1-1&?&   0.1-1&?&  0.1-1&?&  0.1-1&?&  0.1-1&?& $\ll1$ & 5     & 0.5     & 13   & 4       & 10  &  $<0.1$ 
\\[0.1cm] 
Petrol. types\tablefootmark{5}                         &         & 3-5    & & 3-6    & & 3-6   & & 3-6   & & 3-6   & & 1      & 2     & 2       & 3     & 3-6     & 3   & 3 
\\[0.1cm] 
Max. temp.\tablefootmark{5}                            & [K]     & 1020   & & 1220   & & 1220  & & 1220  & & 1220  & & 430    & 670   & 670     & 870   & 1220    & 870
\\
\noalign{\smallskip}
\hline
\end{tabular}
}
\tablefoot{
\tablefoottext{1}{Adapted from \citet{Tri06}, with modifications.} \tablefoottext{2}
{CR group without CH chondrites.} \tablefoottext{3}{From \citet{Bre98}.} \tablefoottext{4}{In matrix.} \tablefoottext{5}{From \citet{Sea88}.} \tablefoottext{6}{From \citet{Con08}.}  \tablefoottext{7}{From \citet{Clo11,Clo11a,Clo12a,Clo12b,Clo12,Clo12c} and \citet{VSc69}.} \tablefoottext{8}{From \citet{Akr98b}.}
A question mark indicates data that are very uncertain.
}

\label{TabMetTypes}
\end{table*}

\section{%
Composition of chondritic material}
\label{SectCompChon}

\subsection{%
Structure of chondrites}
\label{SectCompChond}

The refractory solids of the precursor material from which asteroids formed is composed of two morphologically different components: matrix and chondrules. The matrix is a very fine grained material consisting of particles with diameters $\lesssim1\,\mu$m, while chondrules are glassy rounded beads (sometimes broken) with diameters usually in the range between 0.1\,mm and 1\,mm. In meteorites these two components form a granular material where a significant fraction of the total space filled by this binary mixture are empty pores between the particles. In the precursor material of carbonaceous chondrites part or all of this pore space was once filled with ice.

Table~\ref{TabMetTypes} shows some characteristic data for the different known chondrite groups. The relative abundances of matrix material and chondrules is very different between different groups, ranging for carbonaceous chondrites from almost pure matrix in CI chondrites to a slightly higher volume fraction of chondrules than matrix material in CO chondrites, while for ordinary chondrites the volume fraction of matrix is small to very small, except perhaps for the rare K grouplet. 

The properties of such binary matrix-chondrule mixtures were discussed in \citet{Gai14b}. They can be characterised by the ratio, $f_{\rm ma}$, of matrix volume fraction, $V_{\rm ma}$, to total volume fraction of matrix and chondrules (without the pore space)
\begin{equation}
f_{\rm ma}={V_{\rm ma}\over V_{\rm ma}+V_{\rm ch}}\,.
\label{DefFracMat}
\end{equation}
On average this quantity (see Table~\ref{TabMetTypes}) ranges for ordinary chondrites between $f_{\rm ma}\sim0.1$ and $f_{\rm ma}\sim0.15$ such that the matrix-chondrule mixture in ordinary chondrites of type H, L, LL, EH, and EL is \emph{chondrule dominated} as defined in \cite{Gai14b}, where matrix fills part of the space not occupied by chondrules. In carbonaceous chondrites it varies on average between  $f_{\rm ma}\sim0.42$ and $f_{\rm ma}=1$, such that the matrix-chondrule mixture in carbonaceous chondrites of type CI, CM,and CK is \emph{matrix-dominated},  where chondrules are interspersed within a matrix ground material. The CR, CO and CV chondrites are intermediate cases. 

The chondritic material is porous \citep[see][ for a detailed review]{Con08}. The volume fraction of voids, the porosity
\begin{equation}
\phi={V_{\rm vo}\over V_{\rm ma}+V_{\rm ch}+V_{\rm vo}}\,,
\label{DefPoros}
\end{equation}
with $V_{\rm vo}$ being the volume fraction of void space, varies in meteorites between a few percent and up to about thirty percent \citep{Con08}. Typical values are shown in Table~\ref{TabMetTypes}; for individual meteorites of a group there is significant scatter of $\phi$. These numbers have to be considered with some caution for two reasons. First, shock waves experienced in the course of asteroid collisions, including those  ejecting a meteorite from its parent body, produce cracks in the
material \citep[cf.][]{Con08}. The porosity measured for meteorite specimens is in part impact-generated. In particular low values of $\phi$ can be suspected to be of impact origin. Second, an impact into porous material causes a partial compaction of the target material \citep{Bei14}. The excavated material may show a higher degree of compaction than before the impact. For these reasons the measured porosity of meteoritic material cannot simply be taken as an indigenous property of the parent asteroid material.

The existence of pores in a material obviously has considerable influence on the heat conductivity. For modeling the thermal evolution of a planetesimal one has to reconstruct the initial degree of porosity of the material from which the parent body was formed, because one cannot simply take the porosity measured for meteorites of low petrologic type as the initial state. If one disregards the possible existence of ice in the initial state of the parent bodies of carbonaceous chondrites the initial material is composed of a mixture of chondrules and dust. The diameters of the chondrules vary in principle over a wide range, but most of them are from a rather narrow range of diameters with size ratios typically $\lesssim2$ \citep[e.g.][]{Rub84}. The same most likely holds for the dust particles \citep{Woz12,Woz13}. The random densest packing of granular material of two components with size ratios of less than two deviates only very little from that of a mono-disperse granular medium \c
 itep{Fis94}. We ignore such minor effects and assume for simplicity that the granular units of the chondrule component have all the same diameter $d_{\rm ch}$, and that also the granular units of the dust component all have the same diameter $d_{\rm du}$. 

\begin{table}
\caption{The pure mineral species considered for calculating the properties of
chondrite material, their chemical composition, atomic weight $A$, mass-density $\varrho$ and thermal conductivity at 300 K. They form the indicated solid solutions.}

\begin{tabular}{l@{\hspace{0.2cm}}lrlr}
\hline
\noalign{\smallskip}
Solid solution & chemical & \multicolumn{1}{c}{$A$} & 
   \multicolumn{1}{c}{$\varrho$} & \multicolumn{1}{c}{$K$}  \\
\quad component & formula & & g/cm$^3$ & W/mK \\
\noalign{\smallskip}
\hline
\\
Olivine            &                     &        &      &   \\
\quad forsterite   & Mg$_2$SiO$_4$       & 140.69 & 3.22 & 5.158\tablefootmark{a}  \\
\quad fayalite     & Fe$_2$SiO$_4$       & 203.78 & 4.66 & 3.161\tablefootmark{a}  \\[.1cm]
\multicolumn{2}{l}{Orthopyroxene}        &        &      &   \\
\quad enstatite    & MgSiO$_3$           & 100.39 & 3.20 & 4.961\tablefootmark{a}  \\
\quad ferrosilite  & FeSiO$_3$           & 132.32 & 3.52 & 3.352\tablefootmark{d}   \\[.1cm]
\multicolumn{2}{l}{Clinopyroxene}        &        &      &   \\
\quad wollastonite & Ca$_2$Si$_2$O$_6$   & 116.16 & 2.91 & 4.032\tablefootmark{a} \\
\quad (+ en + fs) \\[.1cm]
\multicolumn{2}{l}{Plagioclase}          &        &      &   \\
\quad anorthite    & CaAl$_2$Si$_2$O$_8$ & 277.41 & 2.75 & 1.679\tablefootmark{a}  \\
\quad albite       & NaAlSi$_3$O$_8$     & 263.02 & 2.63 & 2.349\tablefootmark{a}  \\
\quad orthoclase   & KAlSi$_3$O$_8$      & 278.33 & 2.55 & 2.315\tablefootmark{a}  \\[.1cm]
\multicolumn{2}{l}{Iron-nickel alloy}    &        &      &  \\
\quad iron         & Fe                  &  55.45 & 7.81 & 71.100\tablefootmark{c} \\
\quad nickel       & Ni                  &  58.69 & 8.91 & 80.800\tablefootmark{c} \\[.1cm]
Troilite           & FeS                 &  87.91 & 4.91 & 4.60\tablefootmark{b}   \\
\noalign{\smallskip}
\hline
\end{tabular}
\tablefoot{
\tablefoottext{a}{\citet{Hor69}}, \tablefoottext{b}{\citet{Cla95},} \tablefoottext{c}{\citet{Ho78},} \tablefoottext{d}{extrapolated value from \citet{Hor72}}.
}

\label{TabMinProp}
\end{table}

This two-component mixture is discussed in the appendix of \citet{Gai14b}. It is shown that for such a two component granular medium the porosity in the chondrule dominated mixture is
\begin{equation}
\phi=1-{1-\phi_{\rm ch}\over1-f_{\rm ma}}\,.
\label{AppBinGran}
\end{equation}
Here $\phi_{\rm ch}$ is the porosity of the chondrule component if the matrix would be absent. In the matrix-dominated case it is
\begin{equation}
\phi=1-{1-\phi_{\rm ma}\over1-\phi_{\rm ma}(1-f_{\rm ma})}
\label{AppBinGranMa}
\end{equation}
where $\phi_{\rm ma}$ is the porosity of the matrix component. For a granular medium of equal sized grains there are two critical packings with porosities $\phi\approx0.56$ and $\phi\approx0.64$. The lower value corresponds to the random loose packing which is just stable under weak external pressure, the higher value corresponds to the random densest packing of equal sized spheres \citep[cf. ][]{Jea92}. Since during the growth process of planetesimals the material is joggled again and again it is likely that the porosities $\phi_{\rm ch}$ in the chondrule dominated case and $\phi_{\rm ma}$ in the matrix-dominated case equal the random closest packing that is usually achieved in a granular medium after vigorous tapping and joggling. Table~\ref{TabMetTypes} shows the porosities $\phi_0$ of the granular mixture predicted by such considerations. They are estimates of the likely initial porosity of the material in planetesimals before the onset of sintering. 

For carbonaceous chondrites, except for CK, the highest observed porosities are close to the theoretical values. For ordinary chondrites the predicted value is much lower. This may have a number of reasons, e.g., sintering because of higher peak temperatures or compaction by shocks. The decrease of highest measured porosity of ordinary chondrites with increasing shock stage \citep[Figure 3 of][]{Con08} may be a hint that compaction by impact shock wave played a significant role for observed porosities. For the lowest shock stage 1 the highest observed porosities agree with our model. We take the two-component model as initial structure of chondritic material (corresponding to petrologic type 3).   

For the parent bodies of ordinary chondrites the initially porous chondritic material is compacted by sintering if the body is sufficiently heated by radioactive decays. The pore space vanishes by this and the final structure of the material after compaction corresponds to what is found in meteorites of petrologic type 6. The structure of the compacted material is still complicated because already the initial chondrule and matrix material is a heterogeneous mixture of different minerals and iron. In order to determine the heat conductivity of the material we, thus, have to study how for the initially porous and heterogeneous material the heat conductivity varies from the initial state with decreasing porosity to the final fully compacted but still heterogeneously composed material. This will be accomplished by constructing a numerical model for a granular material with varying compositions of the granular units and by solving numerically the heat conduction equation for this model. 

For the carbonaceous chondrites one has the additional complication that the parent body initially contains ice which, once molten, reacts with the silicates to form phyllosilicates. The corresponding compositional changes (ice bearing $\to$ water bearing $\to$ phyllosilicates) result in changes in the heat conductivity, which have to be included in model computations. Here we concentrate on ordinary chondrites.

\begin{table*} 
\caption{Typical mineral composition of chondrites, mass-densities $\varrho$ of
their components, mass-fractions $X_{\rm min}$ of the components,  their volume fractions $f_{\rm min}$, the heat conductivity $K$ at room temperature of the components, and the resulting bulk value of density and heat conductivity \citep[data from][]{Yom83,VSc69,Jar90}. 
}

\begin{tabular}{l@{\hspace{.7cm}}lcccr@{\hspace{.7cm}}lcccr}
\hline
\noalign{\smallskip}
species & composition & $\varrho$ & $X_{\rm min}$ & $f_{\rm min}$ & \multicolumn{1}{c}{$K$\quad\,}  & composition & $\varrho$ & 
$X_{\rm min}$ & $f_{\rm min}$ & \multicolumn{1}{c}{$K$} \\
        &             & g\,cm$^{-3}$ &  &  &   W/mK & & g\,cm$^{-3}$ & & & W/mK \\
\noalign{\smallskip}
\hline
\noalign{\smallskip}
        & \multicolumn{5}{c}{H-chondrite} & \multicolumn{5}{c}{L-chondrite}  \\ 
\noalign{\smallskip}
Olivine       & Fo$_{80}$Fa$_{20}$ & 3.51 & 0.37 & 0.399 & 4.349 & Fo$_{75}$Fa$_{25}$ &  3.58 & 0.49 & 0.491 & 4.142 \\
Orthopyroxene & En$_{83}$Fs$_{17}$ & 3.25 & 0.25 & 0.291 & 4.150  & En$_{78}$Fs$_{22}$ & 
  3.27 & 0.23 & 0.253 & 3.965 \\
Clinopyroxene & En$_{49}$Fs$_6$Wo$_{45}$ & 3.09 & 0.05 & 0.061 & 4.660 & En$_{48}$Fs$_8$Wo$_{44}$  & 3.10 & 0.06 & 0.070 & 4.660 \\
Plagioclase &  Ab$_{82}$Or$_6$An$_{12}$ & 2.64 & 0.08 & 0.114 & 1.935 &
  Ab$_{84}$Or$_6$An$_{10}$ & 2.64 & 0.08  & 0.109 & 1.985 \\
Nickel-iron & Irn$_{92}$Nkl$_{8}$ & 7.90 & 0.20 & 0.096 & 29.383 & 
  Irn$_{87}$Nkl$_{13}$ & 7.95 & 0.09 & 0.041 & 25.757 \\
Troilite & Tr & 4.91 & 0.05 & 0.039 &  4.600 &  Tr & 4.91 & 0.05 & 0.036 & 4.600 \\
\noalign{\smallskip}
\multicolumn{1}{l}{Bulk value}&  & 3.78 & & & 4.89 & & 3.59 & & & 4.19 \\ 
\noalign{\smallskip}
        & \multicolumn{5}{c}{LL-chondrite} & \multicolumn{5}{c}{EH-chondrite}  \\ 
\noalign{\smallskip}
Olivine       & Fo$_{69}$Fa$_{31}$ & 3.67 & 0.60 & 0.585 & 3.914 & --- &   & &  &  \\
Orthopyroxene & En$_{74}$Fs$_{26}$ & 3.28 & 0.16 & 0.174 & 3.830 & En$_{100}$ & 3.20  & 0.56 & 0.666 & 4.961 \\
Clinopyroxene & En$_{46}$Fs$_{10}$Wo$_{44}$ & 3.10 & 0.05 & 0.058 & 4.660 & --- &  &  &  &  \\
Plagioclase &  Ab$_{86}$Or$_4$An$_{10}$ & 2.64 & 0.09 & 0.122 & 1.985 &
  Ab$_{81}$Or$_4$An$_{15}$ & 2.64 & 0.10 & 0.144 & 1.985 \\
Nickel-iron & Irn$_{70}$Nkl$_{30}$ & 8.14 & 0.04 & 0.018 & 13.636 & 
  Irn$_{92}$Nkl$_{8}$ & 7.90 & 0.25 & 0.120 & 29.383 \\
Troilite & Tr & 4.91 & 0.06 & 0.044 &  4.600 &  Tr & 4.91 & 0.09 & 0.070 & 4.600 \\
\noalign{\smallskip}
\multicolumn{1}{l}{Bulk value}&  & 3.58 & & & 3.78 & & 3.80 & & & 5.62 \\ 
\hline
\end{tabular}

\label{TabMinMix}
\end{table*}

\subsection{%
Mineral components}

The composition of the mineral mixture and the mass fractions, $X_{\rm min}$, or volume fractions, $f_i$, of the components varies between the parent bodies of chondrites. The mass-fractions $X_{\rm min}$ are essential parameters that determine the properties of the material. Because of the marked hierarchy of abundances in the cosmic element mixture, only a few elements can form abundant components in the mixture and a rather limited suite of minerals needs to be considered to determine the gross properties of the material. We consider only the more important components that contain almost the whole inventory of the abundant rock-forming refractory elements (Si, Mg, Fe, Al, Ca, Ni, Na, K). The pure solid substances and their solid solutions that typically are found in a mineral mixture at temperatures below the melting point of the Fe-FeS eutectic ($T=1\,261$\,K,  where differentiation possibly starts, see \citealp{Fei97}) and in the absence of water are given in Table \ref{TabMinProp}. The mixture of minerals in  ordinary chondrites is dominated by these components.

In the applications we consider in particular the mixture of minerals and of the iron-alloy found in H and L chondrites,  taken to be equal to the modal mineralogy of H and L chondrites as given in \citet{Yom83}. This is listed in Table~\ref{TabMinMix}, together with $X_{\rm min}$, the mass-fractions of the different components in the mixture, which are also taken from \citet{Yom83}.

The density of the solid solutions given in the table is simply calculated as average of the densities of the pure components, weighted with their mole fractions in the solid solution. This assumes that non-ideal mixing effects are small and need not to be considered for our calculations.
The average density $\varrho_{\rm bulk}$ of the mixture is calculated from
\begin{equation}
\varrho_{\rm bulk}=\left(\sum_j{X_{{\rm min},j}\over \varrho_j}\right)^{-1}\,.
\end{equation}
This is the density of the consolidated meteoritic material, i.e., without pores. The volume fractions $f_{\rm min,i}$ of the components are
\begin{equation}
f_{\rm min,i}={X_{\rm min,i}\over\varrho_i}\,\varrho_{\rm bulk}\,.
\end{equation}

\begin{figure}

\includegraphics[width=.8\hsize]{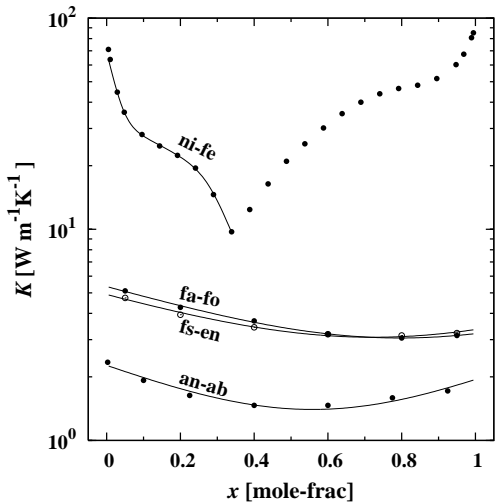}

\caption{Concentration dependence of heat conductivity of the nickel-iron alloy  and of the abundant mineral binary solid solutions (dots). Solid lines correspond to the analytic approximations to the experimental data given in text. The type of solid solution is indicated at the curves. The concentration $x$ of the abscissa refers to the mole fraction of the first component in the solid solution.}

\label{FigVarKX}
\end{figure}

\subsection{%
Thermal conductivity}

This work aims to calculate the heat conductivity $K$ of the mixture of several minerals, metal-alloy, and pores in chondrites from the heat conductivities of the constituents of the mixture. Most of the constituents are not pure substances but are solid solutions with varying concentrations of the solution components in different meteorite classes and with varying heat conductivities depending on the concentrations. Table \ref{TabMinProp} lists the heat conductivities of the pure components at room temperature (300 K) for the pure substances forming the mineral mixture of ordinary chondrites. We refrain for the moment from the temperature dependence of $K$ and consider only the concentration dependencies. The composition of the solid solutions relevant for ordinary chondrites are listed in Table~\ref{TabMinMix}.

Figure \ref{FigVarKX} shows the heat conductivity of the solid solutions of interest. In detail:

\emph{Nickel-iron alloy:} The data are from \citet{Ho78} where the heat conductivity of the nickel-iron alloy is tabulated for a large number of concentrations (mole fractions) between $x=0.0048$ and $x=0.9947$ of the Ni component and for a wide range of temperatures between 4\,K and 1\,100\,K. The figure shows the concentration variation of $K$ at room temperature (300 K). The experimental values can be fitted in the range $x<0.34$ by the following expression
\begin{align}
k_1&=67.863\,{\rm e}^{-17.232\,x}\nonumber\\
k_2&=33.542\,{\rm e}^{-2.0312\,x}\nonumber\\
k_3&=361.44\,{\rm e}^{-10.612\,x}\nonumber\\
K(x)&=\left\{k_1^4+\left(k_2^{-4}+k_3^{-4}\right)^{-1}\right\}^{1/4}
\end{align}
where $x$ is the mole fraction of nickel in the solid solution.  Table~\ref{TabMinMix} shows the interpolated value of $K$ for the alloy in H and L chondrites.

The concentration range of interest for nickel in the nickel-iron alloy in ordinary chondrites is around $x=0.1$, see Table \ref{TabMinMix}. The accuracy of the data for $K$ in this range is $\pm15$\% according to \citet{Ho78}. Additionally the dependence of $K$ on $x$ ist rather strong such that variations of the Ni fraction in individual iron grains in chondrites may result in noticeable variations of $K$. Since we do not consider such variations in our calculations of the effective heat conductivity of the mixed material, the value of the heat conductivity of the Ni,Fe-alloy in chondrites has some uncertainty.   

\emph{Olivine:} This is the solid solution of forsterite and fayalite. The data for heat conductivity of the solid solution at a number of fayalite concentrations shown in Fig.~\ref{FigVarKX} are from \citet{Hor72}. In that paper values are given for a number of composition intervals between the pure end members. The dots in Fig.~\ref{FigVarKX} correspond to the midpoints of the intervals. The data are for room temperature. The experimental values can be fitted by the following expression
\begin{equation}
K(x)=5.3548-5.7704\,x+3.6216\,x^2\,,
\end{equation}
where $x$ is the mole fraction of fayalite in the solid solution.  Table~\ref{TabMinMix} shows the interpolated value of $K$ for the olivine component in H and L chondrites.  

\emph{Orthopyroxene:} This is the solid solution of enstatite and ferrosilite. Data for the concentration dependence of the heat conductivity shown in Fig.~\ref{FigVarKX} are again from \citet{Hor72}, where values for a number of composition intervals between the pure endmembers are given. Only the values for low concentrations of ferrosilite ($x\le0.3$) are experimental data, for higher concentrations they are estimates, but fortunately only data for low Fs concentrations are needed for chondritic material. The data from \citet{Hor72} can be fitted by the following expression
\begin{equation}
K(x)=4.9094-5.0649\,x+3.5078\,x^2\,,
\end{equation}
where $x$ is the mole fraction of ferrosilite in the solid solution.  Table~\ref{TabMinMix} shows the interpolated value of $K$ for the orthopyroxene component in H and L chondrites.  

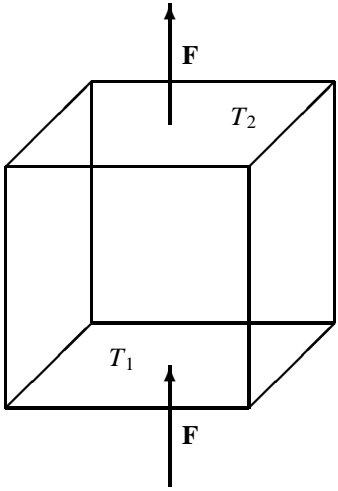
\begin{figure}

\setlength{\unitlength}{.8 cm}
\begin{picture}(8,9)(0,-1)
\put(1,1){\thicklines\line(1,0){4}}
\put(1,5){\thicklines\line(1,0){4}}
\put(1,1){\thicklines\line(0,1){4}}
\put(5,1){\thicklines\line(0,1){4}}
\put(1,5){\thicklines\line(1,1){1.41}}
\put(5,5){\thicklines\line(1,1){1.41}}
\put(1,1){\thicklines\line(1,1){1.41}}
\put(5,1){\thicklines\line(1,1){1.41}}
\put(2.41,2.41){\thicklines\line(1,0){4}}
\put(2.41,6.41){\thicklines\line(1,0){4}}
\put(2.41,2.41){\thicklines\line(0,1){4}}
\put(6.41,2.41){\thicklines\line(0,1){4}}
\put(3.7,5.7){\thicklines\vector(0,1){2}}
\put(3.7,-.3){\thicklines\vector(0,1){2}}
\put(3.9,6.7){$\vec F$}
\put(3.9,0.4){$\vec F$}
\put(2.7,1.7){$T_1$}
\put(4.7,5.7){$T_2$}
\end{picture}

\caption{Cube used for definition of average heat conductivity coefficient.}
\label{FigCube}
\end{figure}

\emph{Plagioclase:} This is the solid solution of albite, anorthite, and a small amount of orthoclase. Data for the heat conductivity of the solid solution of anorthite and albite for different concentration values are given by \citet{Hor72}. No data for the ternary mixture could be found. Since the heat conductivity of pure orthoclase is very similar to that of albite and the orthoclase concentration is small we treat the orthoclase component as albite and calculate the heat conductivity for the modified binary mixture.  The data from \citet{Hor72} for the albite-anorthite solid solution can be fitted by the following expression
\begin{equation}
K(x)=2.2673-3.1000\,x+2.7768\,x^2\,,
\end{equation}
where $x$ is the mole fraction of anorthite in the solid solution.  Table~\ref{TabMinMix} shows the interpolated value of $K$ for the plagioclase component in H and L chondrites.  

\emph{Clinopyroxene:} The composition of the clinopyroxene component in chondrites corresponds to diopside. Table~\ref{TabMinMix} shows the value of $K$ for diopside from \citet{Cla95}.

\emph{Troilite:} The value given in Table~\ref{TabMinMix} is the value for pyrrhotite (Fe$_{1-x}$S) from \citet{Cla95}. A somewhat lower value of $3.53\pm0.05$ is given in \citet{Dim88}. Because troilite is a rare component in the mixture the modification of the effective heat conductivity by using the lower value of $K$ for troilite is negligible.

\section{Heat conductivity of composite materials}
\label{SectHeatCond}

\subsection{Basic model}

In principle it is possible to solve for a composed material such as it is found in meteorites the heat conduction equation 
\begin{equation}
\varrho c_v{\partial\,T\over\partial\,t}=-\nabla\left(K\nabla T\right)
\label{WLG}
\end{equation}
in order to determine the temperature field inside a specimen of that material. Here, $c_v$ is the specific heat capacity of the material, $\varrho$ its mass density, and $K$ the heat conductivity. The spatial variation of the functions $\varrho$, $c_v$, and $K$ is very complicated, corresponding to the distribution of different mineralic  and metallic particles and voids. For solving the heat conduction equation one has to decompose in this case the volume for which the temperature field is to be determined into sub-volumes corresponding to the individual particles of the material and to solve the heat conduction equation inside each particle with the appropriate boundary conditions at the contact areas or at the border to adjacent voids. The corresponding solution will show significant local variations in temperature and heat fluxes, in particular if heat conductivities of the mixture components are significantly different. Such local fluctuations on length scales of the individual
  particle sizes are without interest and it is appropriate to apply a homogenization procedure where one only considers averages performed over regions such that for all relevant mixture components a big number of the individual particles are contained in that region. 

For this purpose we consider for the following an infinitely extended sheet of material of thickness $L$ filled with a multi-component medium and a cube of side-length $L$ within the sheet with two of its surfaces coinciding with the surfaces of the sheet (cf. Fig.~\ref{FigCube}). The two surfaces of the sheet are held at fixed but slightly different temperatures $T_1$ and $T_2$. If two opposite sides of such a body are held at different temperatures one has a net heat flux from the hot to the cold side.  If we solve the heat conduction equation for this cube the heat fluxes through the individual particles of the mixture and their temperature will be slightly different because of different heat conductivities of different components of the material. It can be expected that if the local heat fluxes are averaged over areas with extensions much bigger than the sizes of the biggest particles there results a net heat flux vector that has constant direction and modulus, independent of the
  volume over which the averaging was performed. Also, if one averages the temperature over areas much more extended than the size of the largest particles and perpendicular to the average heat flux vector one will find a constant average temperature over length scales much bigger than the particles and independent of the particular size of the area.

Therefore instead of calculating the heat flux through a composed medium by considering each single particle and the corresponding locally fluctuating heat flux and temperature, we consider only the (locally) averaged temperature distribution and the (locally) averaged heat flux, and an effective heat conductivity of the mixture. This effective heat conductivity is defined as follows: We solve the heat conduction equation for our cube for fixed temperatures at the two front faces and with appropriate boundary conditions at the four cube faces inside the sheet which provides us with the total heat flux through the front faces in the stationary state. Then we define, following the standard way to define a heat conductivity, the average heat conductivity $\langle K_\mathrm{eff}\rangle$ by
\begin{equation}
K_{\rm eff}={L\,F\over T_1-T_2}\,,
\label{DefEffK}
\end{equation}
where $F$ is the heat flux per unit area through the front faces and $L$ the length of the cube edge.

The method how the heat conduction equation is solved numerically is described in Appendix~\ref{NumModHeat}.

\begin{table}[t]

\caption{Effective heat conductivity (units W/mK) at 300\,K of a binary mixture of nickel-iron and forsterite, as determined by solving the heat conduction equation, for different volume fractions $f_{\rm Iro}$ of the iron content and different resolutions $n$. }

\begin{tabular}{lrrlr}
\hline
\hline
\noalign{\smallskip}
comp.  &\multicolumn{3}{c}{numerical} & effective \\
$f_{\rm iro}$  & $n=100$ & $n=200$ & $n=300$ & medium\\
\noalign{\smallskip}
\hline
\noalign{\smallskip}
1.0 & 31.124 & 31.142 & 31.149 & 31.180 \\
0.8 & 22.830 & 22.976 & 23.031 & 24.139 \\
0.6 & 16.286 & 16.409 & 16.457 & 17.589 \\
0.4 & 11.415 & 11.474 & 11.494 & 11.957 \\
0.2 &  7.666 &  7.672 &  7.684 &  7.785 \\
0.0 &  5.187 &  5.188 &  5.188 &  5.188 \\
\noalign{\smallskip}
\hline
\end{tabular}

\label{TabBinOlIr}
\end{table}

\subsection{Two-component mixture of forsterite and nickel-iron}

As one important application the heat conductivity of a non-porous mineral mixture is considered. This kind of structure corresponds to the structure of the compacted material found inside an asteroid for most of its volume and for most time during its thermal evolution. 

The first task for determining an effective heat conductivity of a composite material is to generate a distribution of the different components of the mixture within the cube for which we then solve in a next step the heat conduction equation. The method used to generate a random distribution of components for given volume fractions is described in Appendix~\ref{AppRanDi}.

As a case study we consider a mixture of olivine and nickel-iron. We chose these two components because they are two major components of chondritic material and because their thermal conductivity is considerably different from each other such that the effective conductivity of the mixture is significantly different from that of the components.

\begin{table}[t]

\caption{%
Effective heat conductivity (W/mK) for a mixture with $f_{\rm iro}=0.5$, calculated with resolutions of $n=100$ and $n=200$ for ten different seed numbers $m$ for the random number generator, and average value and average scatter of results. }
 
\begin{tabular}{lllllllllll}
\hline
\hline
\noalign{\smallskip}
 & \multicolumn{5}{c}{$n=100$}\\
\noalign{\smallskip}
$m$ & 1 & 2 & 3 & 4 & 5 \\
$K_{\rm eff}$ & 13.628 & 13.522 & 13.599 & 13.604 & 13.651 \\
$m$ & 6 & 7 & 8 & 9 & 10 \\
$K_{\rm eff}$ &  13.724 & 13.641 & 13.652 & 13.683 & 13.599 \\[.1cm]
average & \multicolumn{5}{l}{$\langle K_{\rm eff}\rangle=13.67\pm0.02$}\\
\noalign{\smallskip}
 & \multicolumn{5}{c}{$n=200$}\\
\noalign{\smallskip}
$m$ & 1 & 2 & 3 & 4 & 5 \\
$K_{\rm eff}$ & 13.702 & 13.597 & 13.672 & 13.673 & 13.737 \\
$m$ & 6 & 7 & 8 & 9 & 10 \\
$K_{\rm eff}$ & 13.816 & 13.718 & 13.707 & 13.773 & 13.7668 \\[.1cm]
average & \multicolumn{5}{l}{$\langle K_{\rm eff}\rangle=13.73\pm0.019$}\\
 
\noalign{\smallskip}
\hline
\end{tabular}
\label{TabBinOlIrRan}

\end{table}

A two-component mixture with volume fractions of the components varying between 0 and 1 in steps of 0.05 are generated by the procedure described in App.~\ref{AppRanDi}. For the heat conductivity $K$ of the components its value at 300 K is used. The temperature at one of the front surfaces of the cube was set to $T_2=310$ K, the temperature at the opposite front surface was set to $T_1=300$\, K. A possible variation of $K$ with temperature was neglected in this model calculation; the temperature dependence is considered later. The cube was decomposed into cells of different sizes by choosing $n=100, 200, 300$. The size of the balls for generating the random mixture is chosen as 1/20 of the cube length $L$. With this ball size the prescribed volume fraction of the mixture components is met by our algorithm with an accuracy better then $6\times10^{-4}$ in all cases. The heat conduction equation was solved for all cases until a stationary temperature distribution has evolved. For this s
 olution, the heat flux through the cube is calculated and via Eq.~(\ref{DefEffK}) the effective heat conductivity $K_{\rm eff}$ is determined.

\begin{figure}[t]

\includegraphics[width=\hsize]{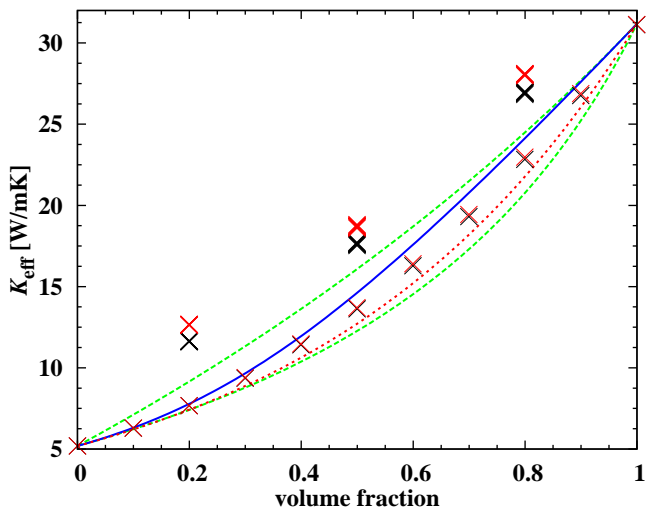}

\caption{Comparison of the effective heat conductivity of a binary mixture of olivine and nickel-iron determined from a solution of the heat conduction equation and compared with predictions of different mixing rules. Crosses: Effective heat conductivity determined from the numerical solution with $n=100$ (black), and $n=200$ (red). The multiple crosses at a volume fraction of 0.2, 0.5, and 0.8, shifted by 3 upwards, correspond to results calculated for different realisations of the random distribution of the mixture components.  Long dashed lines: Theoretical upper and lower limits for heat conductivity of the mixture given by Eqs.~(\ref{KBoundMax}) and (\ref{KBoundMin}), respectively. Short dashed line: Geometric mean according to Eq.~(\ref{HarmMean}). Solid line: Effective heat conductivity according to the Bruggeman effective medium theory, given by Eq.~(\ref{EffMediumBrugg}).}

\label{FigBinOlIr}
\end{figure}

The results are shown in Fig.~\ref{FigBinOlIr} and in Table~\ref{TabBinOlIr} for  resolutions of $n=100$ and $n=200$. As is evident from the figure and the tabular values, the differences between the effective heat conductivities, $K_{\rm eff}$, determined for the solutions with $n=100$ and $n=200$ is less than one percent and the difference between $n=300$ and $n=200$ of the order of 2\textperthousand. This difference is less than the accuracy with which the heat conductivities $K$ of the mixture components are known. This suggests that it suffices to use a resolution of $n=100$ for calculations of the effective heat conductivity for chondritic material as it is needed for model calculations.

Some scatter of the results may be introduced by using different sequences of random numbers for constructing the spatial distribution of the mixture components. Table \ref{TabBinOlIrRan} shows as an example for one case the numerical values of $K_{\rm eff}$ obtained for ten different realisations of the sequences of random numbers using different seed numbers for the random number generator. The average value of the heat conductivity is $\langle K_{\rm eff}\rangle=13.67\pm0.02\,\rm W/mK$ and $\langle K_{\rm eff}\rangle=13.73\pm0.019\,\rm W/mK$ for a resolution of $n=100$ and $n=200$, respectively. The typical relative deviation of the calculated values of a single calculation for the heat conductivity to the  average value of many different calculations using different seed numbers for the random number generator is $\sim0.1$\% for both resolutions. Such a variation is small enough so that it is not necessary to average a large number calculations.

\begin{table}

\caption{Effective heat conductivity (W/mK) at room temperature numerically calculated for the chondritic mineral mixture defined in Table~\ref{TabMinMix} for varying porosity and two resolutions, and effective conductivity calculated with the Bruggeman mixing rule. 
}
 
\begin{tabular}{lllc}
\hline
\hline
\noalign{\smallskip}
porosity   & \multicolumn{2}{c}{Numerical} & Bruggeman \\
           & $n=100$  &  $n=200$ &  \\
\noalign{\smallskip}
\hline
\noalign{\smallskip}
               & \multicolumn{3}{c}{H chondrite} \\
\noalign{\smallskip}
0.00           & $4.870\pm0.012$ & $4.890\pm0.013$ & 4.870 \\
0.05           & $4.460\pm0.007$ & $4.504\pm0.013$ & 4.471 \\
0.10           & $4.072\pm0.018$ & $4.131\pm0.018$ & 4.076 \\
0.15           & $3.684\pm0.011$ & $3.755\pm0.009$ & 3.686 \\
0.20           & $3.319\pm0.020$ & $3.408\pm0.018$ & 3.301 \\
0.25           & $2.957\pm0.028$ & $3.060\pm0.024$ & 2.920 \\
0.30           & $2.608\pm0.026$ & $2.718\pm0.020$ & 2.546 \\  
\noalign{\medskip}
               & \multicolumn{3}{c}{L chondrite} \\
\noalign{\smallskip}
0.00           & $4.186\pm0.005$ & $4.195\pm0.005$ & 4.176 \\
0.05           & $3.835\pm0.008$ & $3.862\pm0.006$ & 3.851 \\
0.10           & $3.502\pm0.015$ & $3.545\pm0.013$ & 3.528 \\
0.15           & $3.170\pm0.007$ & $3.222\pm0.008$ & 3.206 \\
0.20           & $2.856\pm0.018$ & $2.922\pm0.018$ & 2.886 \\
0.25           & $2.552\pm0.021$ & $2.630\pm0.017$ & 2.568 \\
0.30           & $2.255\pm0.022$ & $2.342\pm0.018$ & 2.252 \\  
\noalign{\medskip}
               & \multicolumn{3}{c}{LL chondrite} \\
\noalign{\smallskip}
0.00           & $3.774\pm0.001$ & $3.779\pm0.001$ & 3.770 \\
0.05           & $3.461\pm0.005$ & $3.482\pm0.003$ & 3.484 \\
0.10           & $3.157\pm0.011$ & $3.192\pm0.008$ & 3.199 \\
0.15           & $2.859\pm0.009$ & $2.906\pm0.007$ & 2.914 \\
0.20           & $2.581\pm0.015$ & $2.639\pm0.014$ & 2.629 \\
0.25           & $2.304\pm0.018$ & $2.373\pm0.014$ & 2.345 \\
0.30           & $2.039\pm0.018$ & $2.113\pm0.015$ & 2.061 \\  
\noalign{\medskip}
               & \multicolumn{3}{c}{EH chondrite} \\
\noalign{\smallskip}
0.00           & $5.587\pm0.012$ & $5.621\pm0.012$ & 5.601 \\
0.05           & $5.126\pm0.013$ & $5.189\pm0.013$ & 5.135 \\
0.10           & $4.667\pm0.021$ & $4.739\pm0.021$ & 4.674 \\
0.15           & $4.230\pm0.014$ & $4.320\pm0.016$ & 4.219 \\
0.20           & $3.811\pm0.023$ & $3.915\pm0.024$ & 3.770 \\
0.25           & $3.395\pm0.039$ & $3.513\pm0.035$ & 3.328 \\
0.30           & $2.992\pm0.028$ & $3.126\pm0.028$ & 2.893 \\  
\noalign{\smallskip}
\hline
\end{tabular}

\label{TabKeffChond}
\end{table}

The differences resulting for different realisations of the distribution of the components with prescribed volume fractions are small such that there is no need to determine averages for a large collection of realisations. Just a few ones suffice.  

\begin{figure*}
\sidecaption
\includegraphics[width=12cm]{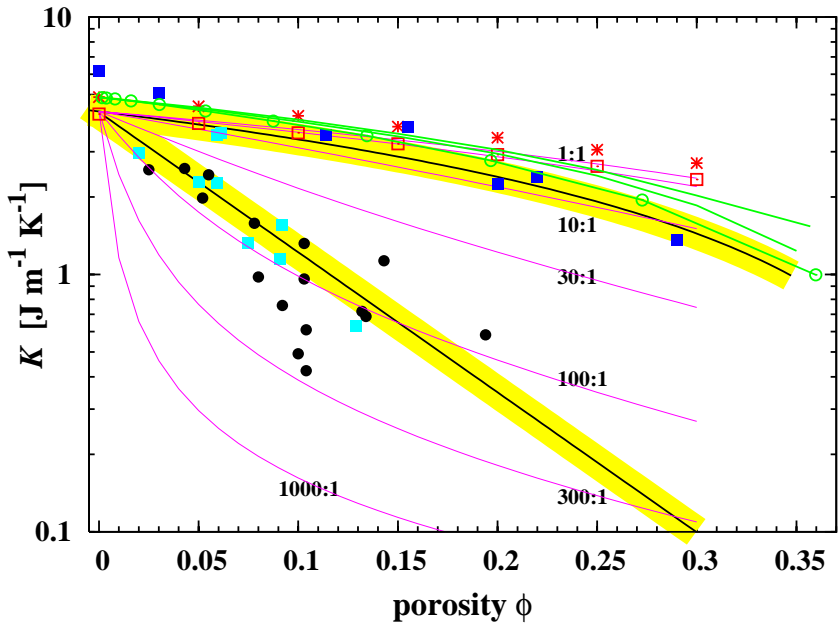}
\caption{%
Comparison of calculated heat conductivities of porous chondritic material and measured data given in \citet{Yom83} with some additional data from \citet{Ope12}.  Lower solid black line: Least square fit to  data (filled cyan squares: H chondrites, filled circles: L chondrites). 
Red symbols: Results for a numerical simulation including some pore space corresponding to  porosity $\phi$ (crosses: H chondrite, open squares: L chondrite material). The blue filled squares are measured values for sandstone \citep{Des74}. Solid green lines: Simulated sintered granular material. Solid green line with circles: Sintered granular material of glass beads described in Sect.~\ref{SectSimGranSint}. The lilac lines are for crack-like voids simulated by the Bruggeman mixing rule with rotational ellipsoid of the indicated axis ratio as inclusions. Upper solid black line: approximation (\ref{ApprKphiGr0}) for sintered material. The two broad strips highlight the differences of the porosity variations of $K$ for impact-cracked and sintered granular material.
}

\label{FigKChond}
\end{figure*}

\subsection{%
Analytical approximations for heterogeneous media}

For technical applications a number of semi-empirical mixing rules have been developed to calculate the heat conductivity of a composite material from the properties of its components. An overview on the problem can be found, e.g., in \citet{Ber95}. In the following we compare some of these approximations with the results of our numerical calculations in order to check their applicability to thermal models of asteroids.
 
One simple approximation is the geometric mean \citep[see][]{Cla95}
\begin{align}
K_{\rm eff,g}&=\prod_{i=1}^N\,\bigl(K_i)^ {f_i}\,,
\label{HarmMean}
\end{align}
where $f_i$ are the volume fractions of the components and $K_i$ the individual heat conductivities of the components. The geometric mean is frequently applied in numerical model calculations to determine the heat conductivity of a mixture. Figure \ref{FigBinOlIr} shows the effective heat conductivity calculated from this approximation for the olivine-iron mixture as short-dashed line. The deviation between this and the numerically calculated effective heat conductivity of the mixture is significant. For the mixture here considered the geometric mean is only an approximation of low accuracy that should be avoided. 

We define, following \citet{Ber95}, the function
\begin{equation}
\Sigma(z)= \left( \sum^{N}_{i=1} \frac{f_i}{2z+K_i} \right)^{-1}-2z\,.
\label{GenMix}
\end{equation}
An upper and lower bound for the heat conductivity of the mixture is given by \citep[cf.][]{Ber95}
\begin{align}
K_{\rm eff, max}&=\Sigma(K_{\rm max})\,,\label{KBoundMax}\\
K_{\rm eff, min}&=\Sigma(K_{\rm min})\label{KBoundMin}
\end{align}
where $K_{\rm min}$ and  $K_{\rm max}$ are the lowest and highest values of the heat conductivities of the mixture components, respectively. Figure \ref{FigBinOlIr} shows both limits to $K_{\rm eff}$ as long-dashed lines. The computed values of $K_{\rm eff}$ are between these limits.

Another approach are effective medium theories \citep[cf.][]{Ber95} resulting in certain mixing rules. A frequently used rule is the Bruggeman mixing rule \citep{Bru35} which follows from Eq.~(\ref{GenMix}) by letting $\Sigma(z)=K_{\rm eff,B}$ be the effective heat conductivity. This defines the effective heat conductivity $K_{\rm eff,B}$ as the solution of the equation
\begin{equation}
K_{\rm eff,B} = \left( \sum^{N}_{i=1} \frac{3f_i}{2K_{\rm eff,B} + K_i} \right)^{-1}\,.
\label{EffMediumBrugg}
\end{equation}
In Fig.~\ref{FigBinOlIr} the effective heat conductivity calculated from this equation is shown as a solid line. For our problem the effective heat conductivity predicted from the Bruggeman mixing rule is rather close to what is found from solving numerically the heat conduction equation.  An inspection of Table~\ref{TabBinOlIr} confirms this. This suggests that for compacted chondritic material the effective heat conductivity can be determined with reasonable accuracy from Eq.~(\ref{EffMediumBrugg}).

\subsection{%
Chondritic mixtures}

Next the effective heat conductivity is calculated for the six-component mixtures of ordinary chondrites as defined in Table~\ref{TabMinMix}. The numerical calculation is done for resolutions of $n=100$ and $n=200$ and for ten different seed numbers for the random number generater in order to obtain different realisations of the mixture with prescibed average composition. The results are shown in Table~\ref{TabKeffChond} (and also in Table~\ref{TabMinMix}). 

The results for vanishing porosity $\phi=0$ correspond to a completely compacted material, e.g., as it is encountered in chondrites of petrologic type 6. This cannot be compared directly to laboratory determined values for the heat conductivity of chondrites because the material of chondrites is always somewhat porous  because they were subject to impact generated shocks which generate fractures in the material even if it was initially completely compacted. Therefore we have to extrapolate the measured data to zero porosity. Despite strong scatter the data of \citet{Yom83} show a clear systematic trend of the heat conductivity with porosity (cf. Fig.~\ref{FigKChond}), which was used in \citet{Hen11} to fit the data for H and L chondrites with an analytic fit formula. Unfortunately the number of data is too small and their scatter too high as to obtain separate fits for H and L chondrites. The average heat conductivity was fitted by
\begin{equation}
K=4.3\,{\rm e}^{-\phi/0.08}\,\rm W\,m^{-1}K^{-1}\,.
\label{ApprKphiMet}
\end{equation}
The average heat conductivity extrapolated to zero porosity is $K_0=4.3$\,W\,m$^{-1}$K$^{-1}$. The uncertainty of the least square fit for the value of $K_0$ is $\pm 0.67$\,W\,m$^{-1}$K$^{-1}$. In view of the paucity and strong scatter of available data the agreement between the extrapolated value of measured heat conductivities to zero porosity and the results of the theoretical calculation of $K=4.89$\,W\,m$^{-1}$K$^{-1}$ for H chondrites and $K=4.19$\,W\,m$^{-1}$K$^{-1}$ for L chondrites can be considered as satisfactory. Therefore it is possible to reliably predict the average heat conductivity of the mineral mixture of chondrites by our numerical model. 
 
The numerical results are compared in Table~\ref{TabKeffChond} also with results for the effective heat conductivities calculated by the Bruggeman mixing rule, Eq.~(\ref{EffMediumBrugg}). The result of the Bruggeman mixing rule is close to the numerically calculated value. Since the accuracy of the input data for the heat conductivity of the pure components is low  \citep[cf. the tables in][]{Cla95}, the accuracy of the mixing rule probably suffices in practical applications to calculate the average heat conductivity of a multi-component medium.


\section{Heat conductivity of porous material}
\label{SectGran}

In the following we study the dependence of the heat conductivity of a granular material on the porosity.

\subsection{%
Porous material with randomly distributed pores}

A porosity of the chondritic material is simulated by introducing a new component in the mixture with a heat conductivity coefficient of $K=0.01$. This value is sufficiently small such that the volume elements filled with this fictitious component can be considered as effectively insulating in comparison with the more than a factor of hundred times higher conductivity of the other components. On the other hand, the jump in the value of $K$ between adjacent cells of different conductivity is sufficiently small that the finite difference solution method still works and no special treatment at internal interfaces to vacuum is required.
   
The first problem we consider is a porous material generated by the method described above. This results for low porosities in voids interspersed in an otherwise compact material. This can be assumed to describe the situation encountered in a sintered material shortly before and after closure of the pore space. 

Table \ref{TabKeffChond} shows the results of our calculations for the average heat conductivity at different porosities. For comparison, the results obtained by the Bruggeman mixing rule are also shown. The Bruggeman mixing rule yields results that agree reasonable well with the numerical results and may also be used for porous material.

We compare in Fig.~\ref{FigKChond} the results of our numerical model with measurements of the heat conductivity of sandstone \citep[data from][]{Des74}. Sandstone is chosen because this is a material that is formed by compaction of a granular material. In this respect one could expect a close similarity between our model and sintered chondritic material with sandstone, except that the granules in the latter case are mainly quartz particles. The data points prom \citet{Des74} for $\phi>0$ correspond in most cases to a material with a volume fraction of $>95$\% quartz and some admixture of other minerals (e.g., feldspar). The data point with $\phi=0$ corresponds to pure quartz \citep[from][]{Cla95}. This is not the same mineral composition as in chondrites, but the heat conductivity of the compact material is with a value of 6.15\,W\,K$^{-1}$m$^{-1}$ for the compact material very similar to chondritic material such that the variation with porosity can be directly compared.   

An inspection of Fig.~\ref{FigKChond} shows that our numerical model predicts a variation of $K$ with $\phi$ which is very similar with what is observed for sandstone. Only the decrease of heat conductivity with increasing porosity is slightly steeper for sandstone as in our model. Hence, the numerical model with a random distribution of pores is able to reproduce the heat conduction properties of a compacted material. There is only a slight deviation at increasing porosity which may be attributed to different microscopic structures between our random distribution of the components and the decisively granular structure of sandstone. This is considered in the next section.  

\begin{figure}[t]

\includegraphics[width=.49\hsize]{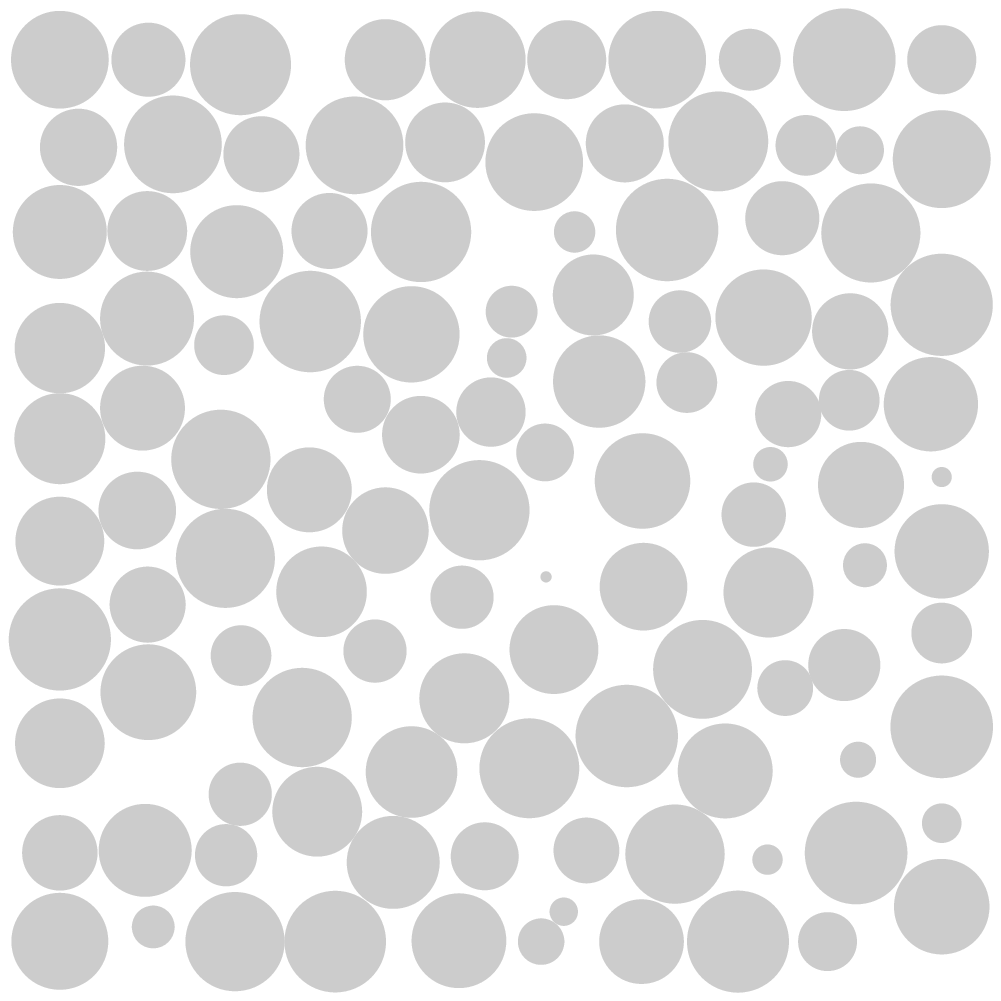}
\hfill
\includegraphics[width=.49\hsize]{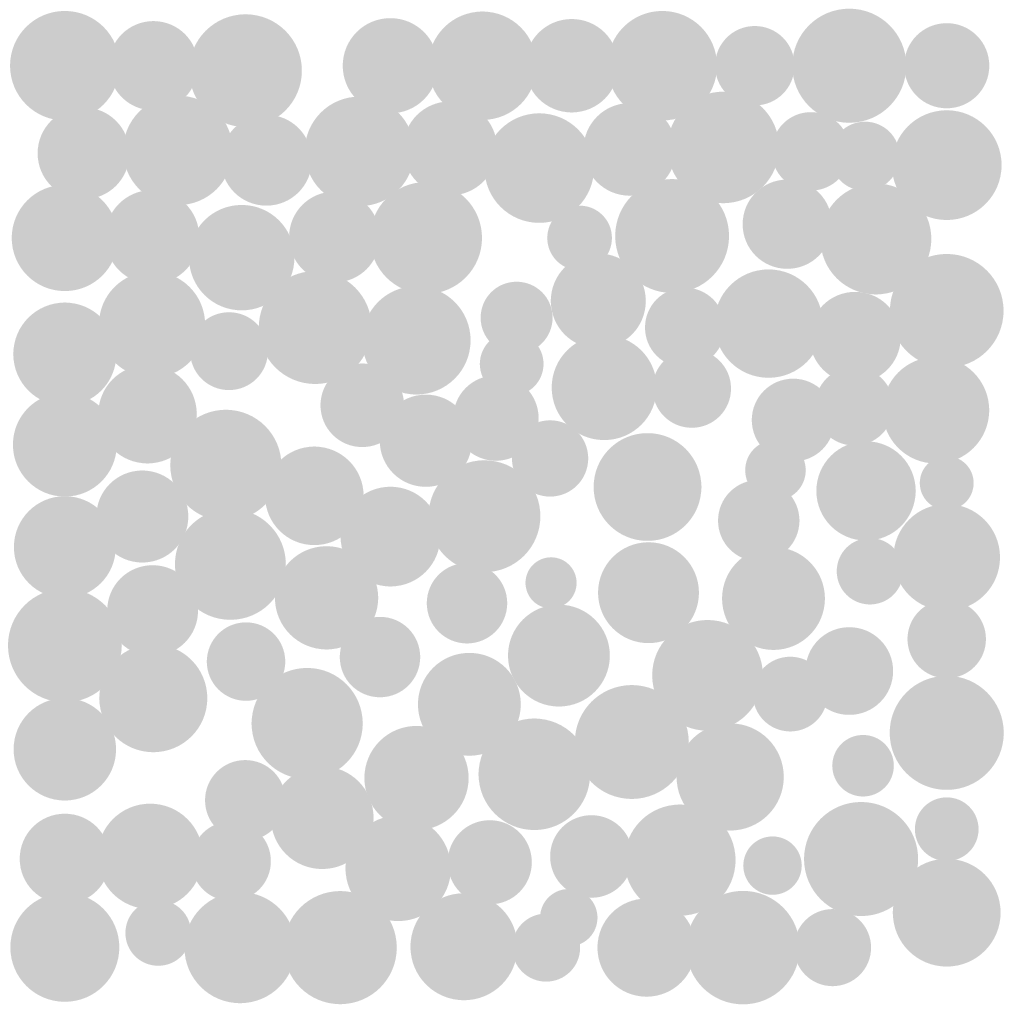}

\caption{Cut along the midplane orthogonal to one of the coordinate axes through a box filled with a random packing of equal sized spheres. The filled circles correspond to the interior of a ball. \emph{Left:} Before sintering. \emph{Right:} With simulated sintering.}

\label{FigRandPackSph}
\end{figure}

\subsection{Granular material}
\label{SectSimGranSint}

For studying a material that has a structure more close to chondritic meteorites or sandstone, we construct a numerical model for a granular medium formed by a packing of balls. For this purpose we simulate the following process: In a box filled with a viscous medium a large number of balls with prescribed distribution of radii is inserted at the top at random positions (generated with a random number generator). The balls fall under the action of a downward directed force. The balls interact with a repulsive force with the wall and mutually with each other if the distance of the centre of a ball to a wall becomes smaller than the radius of the ball or if the distance between the centres of two balls becomes smaller than the sum of their radii, respectively. The bottom of the box and one of the sidewalls (this suffices) can be in vibrational motion for some period of time to simulate joggling of the box. The equations of motions of the balls are solved by a simple numerical procedure
 ; the details of the model are described in Appendix~\ref{NumModGran}.

After switching off the vibrations the balls settle to a random packing of spheres because of the frictional force. As an example, the random packing of 2800 equal sized spheres in a box obtained by this procedure is shown in Fig.~\ref{FigRandPackSph}. It is also possible to use some distribution of radii, but for the moment we consider this somewhat simplified model for the chondrules existing in the parent body of chondritic meteorites before the onset of sintering. In order to get rid of wall effects where the packing fraction is lower than in an unlimited medium, we cut out from our packing the central region by omitting about two layers adjacent to the wall. The porosity of the arrangement of balls obtained in this way is found to be 0.35, i.e., the packing corresponds to the random close packing of balls \citep[cf.][]{Gai14b}. This packing of balls can be considered as a simple model for the packing of chondrules in chondritic material before sintering.

One can homologous shrink this package by reducing the coordinates of the centres of the balls with respect to the centre of gravity by a fixed factor for all three directions of space. This generates a certain degree of overlap between the balls and correspondingly reduces the total void volume. This is just the way how it is assumed in the model of \citet[ see also the discussion in  \citealt{Hen11}]{Arz83} that sintering of a package of balls can be modelled, except that here the ball radii are held fixed and ball distances are scaled while in the sintering model ball radii and distances both are scaled. In this way one can simulate the sintering of the ensemble of balls and generate numerical models for different stages of the sintering process. 

\begin{table}

\caption{%
Average heat conductivity (W\,K$^{-1}$m$^{-1}$)for random packing of balls (chondrules) with different porosities $\phi$ of some given material, normalized to the heat conductivity of the compact material ($\phi=0$). Model 1: Closest packing of equal sized spheres. Model 2: Packing of spheres with two different radii in ratio 2:1 and equal abundances. Model 3: Packing of spheres with a distribution of sizes around some mean (see text). Model 4: Body of sintered glass-beads.}

\begin{tabular}{llllllll}
\hline
\hline
\noalign{\smallskip}
\multicolumn{2}{c}{model 1} & \multicolumn{2}{c}{model 2} & \multicolumn{2}{c}{model 3} & \multicolumn{2}{c}{model 4} \\
$\phi$ & $K$ \\
\noalign{\smallskip}
\hline
\noalign{\smallskip}
0.38 & 0.315 & 0.35 & 0.253 & 0.35 & 0.290 & 0.359 & 0.204 \\
0.30 & 0.415 & 0.30 & 0.380 & 0.30 & 0.418 & 0.273 & 0.399 \\
0.25 & 0.522 & 0.25 & 0.496 & 0.25 & 0.526 & 0.197 & 0.567 \\
0.20 & 0.625 & 0.20 & 0.604 & 0.20 & 0.625 & 0.134 & 0.709 \\
0.15 & 0.722 & 0.15 & 0.705 & 0.15 & 0.721 & 0.087 & 0.808 \\
0.10 & 0.819 & 0.10 & 0.804 & 0.10 & 0.817 & 0.053 & 0.884 \\
0.05 & 0.912 & 0.05 & 0.905 & 0.05 & 0.912 & 0.030 & 0.937 \\
0.00 & 1.000 & 0.00 & 1.000 & 0.00 & 1.000 & 0.008 & 0.985 \\
\noalign{\smallskip}
\hline
\end{tabular}

\label{TabKBalls}
\end{table}

We decomposed the ball packing into a grid of cubes by counting a cube as belonging to a ball if for at least one ball the midpoint of the cube is located inside a ball. The remaining cubes are counted as void space. We calculated the effective heat conductivity of such a model by assuming for the ball material the heat conductivity of H chondrite material from Table \ref{TabKeffChond} and assigning to the void space artificially a small conductivity as in the preceding model. Figure~\ref{FigKChond} and model 1 from Table \ref{TabKBalls} show the resulting variation of $K$ with $\phi$ for this model. As a second case a packing of 3\,800 balls with two different radii with ratio 2:1 is also shown in Fig.~\ref{FigKChond} and by model 2 in Table~\ref{TabKBalls}. The bigger number of balls is required in this case to fill about the same volume. We obtain now an improved representation of the porosity dependence of the average heat conductivity of sandstone material with voids filled by a
 ir. Thus, the numerical method used here is suited to handle such type of porous materials.
 
We consider also an empirically determined size distribution of H chondrites as given in \citet[ which give also distributions for other meteorite classes]{Fri14}. The data are taken from their Fig.~1 for Hammond Downs (H4). The distribution of chondrule diameters is peaked around a diameter of 0.46\,mm with halfwidth of $\sim\pm0.15$\,mm. The results for the heat conductivity calculated for such a mixture of spheres is shown in Table~\ref{TabKBalls} as model 3. The results are closer to that of the mono-dispersed case (model 1) than to the binary mixture with a 2:1 size ratio (model 2). In view of the  relatively sharply peaked size distribution of the chondrules in H chondrites this was to be expected.

Finally we obtained from F. M\"ohlmann, E. Beitz, and J. Blum  (working group planetary formation at the institute for geophysics and extraterrestrial physics at the university of Braunschweig, Germany) data for a sintered cube of originally equal sized glass beads, for which the spatial structure was determined by tomography, resulting in a data file for the decomposition of the body into a grid of cubes. For this, packings of different porosities were generated by shrinking or extending the original spheres. The results of the calculation of the heat conductivity of such granular material are shown in Table~\ref{TabKBalls} as model 4 and in Fig.~\ref{FigKChond} as green line with circles. The results are comparable to the other cases considered and the variation of $K$ with $\phi$ rather closely follows that of natural sandstone. 

The theoretical modelling of heat conductivity of sintered granular material with spheres as primitive granular units and the natural specimens of sandstone show a high conformity in the conductivity variation with porosity. This obviously results from the similarity of the geometric structure of the void space in such materials. Since the material of ordinary chondrites is dominated by the almost spherical chondrules it is obvious to assume that the variation of heat conductivity of chondritic material with porosity during sintering in an up-heating parent body is the same as it is found in this calculation.   
 
\subsubsection{%
Approximation for heat conductivity}

For the purpose of model calculations it is advantageous to use an analytic expression for the porosity dependence of the heat conductivity. The variation of our results for the mean heat conductivity of partially sintered granular material with varying porosity can be approximated by 
\begin{equation}
K(\phi)=K_\mathrm{b}\left(1-2.216\,\phi\right)\,,
\label{ApprKphiGr0}
\end{equation}
where $K_\mathrm{b}$ is the bulk heat conductivity of the compact material. This expression goes to zero for $\phi=0.451$ such that we better clip negative values by letting
\begin{equation}
K_1(\phi)=K_\mathrm{b}\max\left(\left(1-2.216\,\phi\right), 0\right)\,.
\label{ApprKphiGr}
\end{equation}
A zero heat conductivity is unrealistic, however. For high porosity the heat conductivity should tend to the heat conductivity of highly porous material as studied, e.g., in \cite{Kra11}. We have approximated their results for very porous materials by
\begin{equation}
K_2(\phi)=K_\mathrm{b}\,{\rm e}^{1.2-\phi/0.167}\,.
\label{ApprKphiFluff}
\end{equation}
As described in that papers we join both approximations into a single prescription for calculating $K(\phi)$ in the form
\begin{equation}
K(\phi)=\left((K_1^4(\phi)+K_2^4(\phi)\right)^{1\over4}\,.
\label{IntKpor}
\end{equation}
This interpolates smoothly between the limit cases of low and high porosity. The bulk heat conductivity $K_\mathrm{b}$ is calculated for composed material as described in this work.

\subsection{%
Comparison with meteoritic data}

The results of the model calculations for the heat conductivity of chondritic material are compared in Fig.~\ref{FigKChond} with meteoritic data of some H and K chondrites  \citep{Yom83}. The dependence of the heat conductivity on porosity found in our numerical models agrees rather well with the experimental results for the case of sandstone with voids filled by air. Our results strongly deviate, however, from the experimental results for meteorites. Obviously there is a strong discrepancy between the numerical results for our model of a porous material and the empirical data. This hints to a fundamental difference between the structure of void space in compacted spherical granules and the void space that is encountered in chondritic material.  

In the material which we considered the voids are interstitials between some kind of granular units -- however the structure of void space in the meteoritic material is different. For meteorites, even if the shock stage is low, the material contains numerous micro-cracks which are mainly responsible for the observed porosity of ordinary chondrites \citep{Con08}. They result from impact induced shock waves that traversed the meteoritic material and fractured the material in a certain volume around the impact site. These cracks are essentially planar structures which act as efficient barriers for the heat flux, even if the corresponding void volume associated with them is small. The most plausible explanation for the different dependencies on $\phi$ in materials with a void-space structure like that in sandstone on the onehand side and like that for meteoritic material on the otherhand side seems to be the strongly different geometry of the voids in both cases. This also means that the
  pore space measured for meteorites is in large part not pristine but a secondary product of later bombardment of the parent body with boulders since the formation of the body. 

\subsubsection{%
Crack-like voids}

In order to check the hypothesis that the experimental values obtained for meteorites reflect the large number of cracks we model such cracks by insulating inclusions with heat conductivity $K_\mathrm{i}$ and with a shape corresponding to strongly flattened rotational ellipsoids, embedded in an otherwise homogeneous material with heat conductivity $K_\mathrm{m}$. For such a model the effective heat conductivity $K_\mathrm{eff}$ can be calculated from the Bruggeman mixing rule for non-spherical inclusions \citep[cf.][]{Ber95} 
\begin{equation}
\sum\limits_{n=1}^Nf_n\left(K_n-K_\mathrm{eff}\right){1\over9}\sum_{j=1}^3{1\over L_jK_n+(1-L_j)K_\mathrm{eff}}=0\,,
\end{equation}
where the sum $n$ runs over all components in the mixture (inclusions and matrix in our case), $f_n$ is their volume fraction and $K_n$ their heat conductivity, the sum $j$ runs over the principal axis of the ellipsoid. This is a non-linear equation for $K_\mathrm{eff}$ which has to be solved numerically. The $L_j$ are the depolarisation coefficients. These are given for an oblate rotational ellipsoid with principal axis $A=B>C$ by \citep{Boh83}
\begin{equation}
L_1={g\over2e^2}\left({\pi\over2}-\arctan g\right)-{g^2\over2}\,, 
\end{equation}
with $L_2=L_1$ and $L_1+L_2+L_3=1$, where
\begin{equation}
e^2=1-{C^2/A^2}\,,\quad g={\sqrt{1-e^2}\over e}\,.
\end{equation}
We assume for the matrix $K_{\rm m}=4.3$ W\,m$^{-1}$K$^{-1}$ as derived by extrapolation  to zero porosity for the chondrites considered in \citet{Yom83} and $K_i=10^{-4}$  W\,m$^{-1}$K$^{-1}$ for the insulating inclusions, and we vary the axis ratio from $A:C=1:1$ to $A:C=1000:1$. The resulting run of $K_\mathrm{eff}$ with porosity is shown in Fig.~\ref{FigKChond} as the lilac lines. Most experimental values are within a factor of about two near the line with $A:C=100:1$. This suggests that, indeed, the porosity variation of the heat conductivity observed for chondritic material is determined by crack-like voids in the meteorites.   


\section{%
Application to H chondrite parent body}
\label{SectParBod}

\subsection{Porosity of surface layers}

\label{SectSurfPorEvol}

According to our present knowledge planetesimals form by some process that assembles free floating dust and chondrules from the accretion disk into planetesimals. The details of this process are not important for our problem, only the fact that initially the material should be a loosely packed bimodal granular medium consisting of about mm-sized granules and less than $\mu$m sized dust. 

The growth of bodies to the size of the parent bodies of meteorites involves according to present models low relative velocities of $\lesssim1\,\rm km\,s^{-1}$ \citep[see][for a review on asteroid formation]{Joh14,Joh15}. These are much too low to shatter the dust particles and chondrules. It is only if Jupiter has formed that relative velocities in the region of the asteroid belt are pumped up to the order of $5\,\rm km\,s^{-1}$ where collisions become destructive. At that time obviously bodies with sizes up to several hundred km had already formed from which the parent bodies of the chondrites are the survivers. The initial structure of the material from which the bodies are built, thus, is generally assumed to  be a granular material which is to some extent compressed by the action of self-gravitation of the body and by collisions during the growth process.  

Heating by decay of radioactives raises the temperature within less than 1 Ma after formation above the critical temperature of $\sim900$\,K where sintering removes the pores from the material \citep{Gai14b} resulting in compact rocky material. Under certain circumstances the central temperature may become even high enough for melt migration of iron such that they become partially differentiated \citep{Elk11}. However, there remains an outer layer where temperatures remain insufficient for sintering. This layer preserves the initial granular structure. Such incompletely compacted but thermally metamorphosed material has been found to form the material of a number of ordinary chondrites and has been studied by micro-tomography in some detail by \citet{Fri08,Sas09,Fri13,Fri14b}. Such chondrites have large porosities and most of their pore space is intergranular; abundant micro-cracks are not found. A detailed discussion of such chondrites is given in \citet{Fri14b} who argued that they
  are ``rare materials that survived the earliest stages of impact processing on ordinary chondrite parent bodies''. The existence of such material shows that the initial texture of the material of the parent bodies resembles the structure of granular material as studied in the preceding section.  
  
After formation of Jupiter (at some still unknown instant, but probably later than 2\,Ma after CAIs) the relative velocities of bodies in the asteroid belt are stirred up and hypervelocity impacts began which result in cratering of the surface of the bodies and formation of an outer regolith layer. The formation of the regolith layer is obviously accompanied by extensive formation of cracks in the material. If one models the thermal evolution of the parent bodies of meteorites, one has to discriminate between the structure of the void space of the initial sand-like mixture of chondrules and matrix from which the body formed and the void space structure induced in material compacted by impacts. The heat conductivity and the way how it depends on void space is very different in both cases.

It is not possible to use the values measured for meteorites for model calculations of the \emph{early} thermal evolution of meteoritic parent bodies, as it has been done up to now in our previous papers and by others \citep{Yom84,Ben96,Akr98,Har10, Hen11, Hen12,Hen13}. For the earliest phase of evolution one has to use instead the results for compacted porous material for which sandstone seems to be an analogue. Later the properties of the surface material are modified by impacts and abundant cracks are formed. It is then necessary to consider how the properties of the outer few km of the body develop under the action of a bombardment from outside. In particular it is important how the time to form an impact caused regolith layer is related to the characteristic heating and cooling time of the outer layers.   

The formation of a regolith layer on the surface of asteroidal bodies has been studied by \citet{Hou79,War02,War11}. Explicit results for the time required to build-up such a layer are only given in the paper by \citet{Hou79}. They considered impacts into basaltic and into sand-like material, but for bodies of 100 km and 300 km diameter only results for impacts into basaltic material are given. From their figures one infers that the formation of a full-fledged regolith layer lasts 1 \dots\ 2 Ga. This is significantly longer than the about $\sim100$ Ma period for the essential part of the thermal evolution of the parent bodies. During this initial period only a thin regolith cover of the surface forms which means within the frame of their model that only some fraction of the surface suffered cratering by impacts. Since their model deals with average values, the real covering of the surface by craters and their ejecta during the initial growth phase should be rather patchy and large pa
 rts of the surface are likely not yet affected \citep[cf. also][ Fig. 11]{War02}. By a simple test calculation we show later (see Fig.~\ref{FigTestMod}) that a thin early regolith layer does not strongly affect the initial thermal evolution of a parent body and can be neglected in zero order approximation. But it is clear that the porosity evolution by impacts should be included in future model calculations.

\begin{table}

\caption{Basic parameters used for the sample models}

\begin{tabular}{lllll}
\hline
\hline
\noalign{\smallskip}
Quantity & Symbol & Value & Unit \\
\noalign{\smallskip}
\hline
\noalign{\smallskip}
Radius chondr. & $G$ & 150 & $\mu$m \\
Initial porosity & $\phi_0$ & 0.248 \\
Bulk heat cond. & $K_\mathrm{b}$ & 4.9 & W/mK \\
$^{26\!}$Al/$^{27\!}$Al ratio &  &$5.1\times10^{-5}$ & \\
$^{60}$Fe/$^{56}$Fe ratio & & $1.15\times10^{-8}$ & \\[.3cm]
Radius & $R_\mathrm{p}$ & 150 & km \\
Formation time & $t_\mathrm{form}$ & 2.0 & Ma \\
Surface Temp. & $T_\mathrm{srf}$ & 200 & K \\
Surface pressure & $p_{\rm srf}$    & 10  & Pa\\
\noalign{\smallskip}
\hline
\end{tabular}

\label{TabParmTest}
\end{table}

\subsection{Evolution model}

\subsubsection{Model calculation}

Our results for the heat conductivity of the chondritic material are used to calculate thermal evolution models of pla\-netesimals. The method for the model construction is described in \citet{Hen11}. The numerical method for solving the heat conduction equation is now changed from a method based on finite differences to a method based on finite elements. The improved treatment of sintering described in \citet{Gai14b} is included. The instantaneous formation approximation is used since it was found that this fits best to the observed cooling properties of H chondrites \citep{Hen13}. Additionally melting of the Fe,Ni-FeS complex and of silicates is implemented (the method for modelling will be described elsewhere), but no melt migration is considered. Melting is included in the modelling because the parent body of ordinary chondrites could have been partially molten in their innermost core regions, though presently no chondrites are known which show indications of partial melting and 
 at the same time seem to originate from one of the parent bodies of the ordinary chondrites.

\begin{figure*}
\sidecaption
\includegraphics[width=12cm]{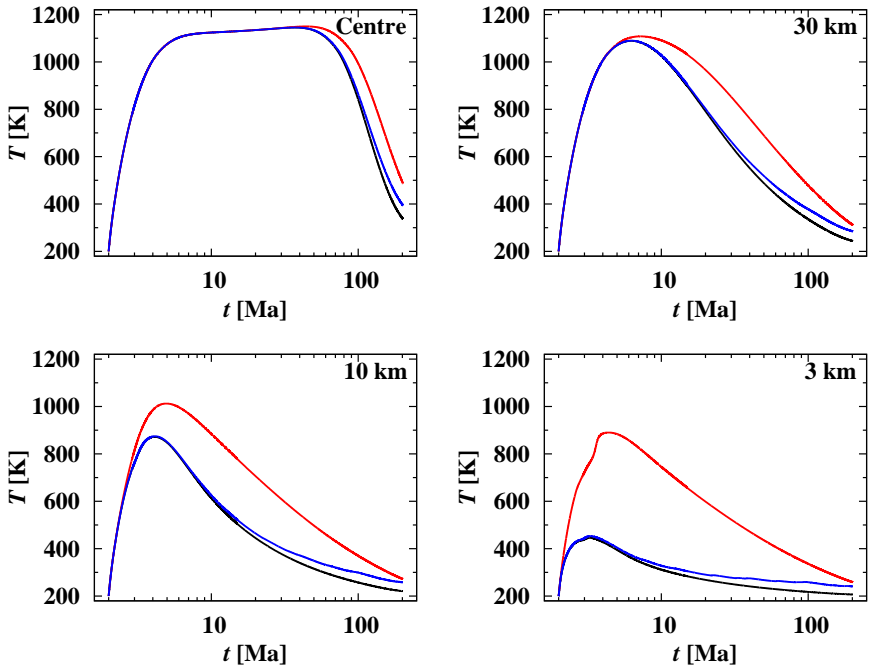}
\caption{%
Comparison of the temperature evolution of a parent body at depth of 3, 10, and 30 km and at the centre of the body with different assumptions with respect to the porosity of the surface layers. Red lines: model Y. Black lines: Model G. Blue lines: Model with growing outer regolith layer.}

\label{FigTestMod}
\end{figure*}

The basic parameter values used for the following model calculations are listed in Table~\ref{TabParmTest}. We consider parent bodies of ordinary chondrites where chondrules strongly dominate over matrix material. The basic properties of this mixture are listed in Table~\ref{TabMetTypes}. It is assumed that the properties of the initial chondrule matrix mixture do not vary within the body. The initial porosity is calculated as described in Sect.~\ref{SectCompChond} (see Table~\ref{TabMetTypes}). The sinter properties of the material of ordinary chondrites are dominated by the chondrules. The method for calculating the sintering of such material and the set of parameters describing the properties of the material are described in \citet{Gai14b}.

\begin{figure}

\includegraphics[width=\hsize]{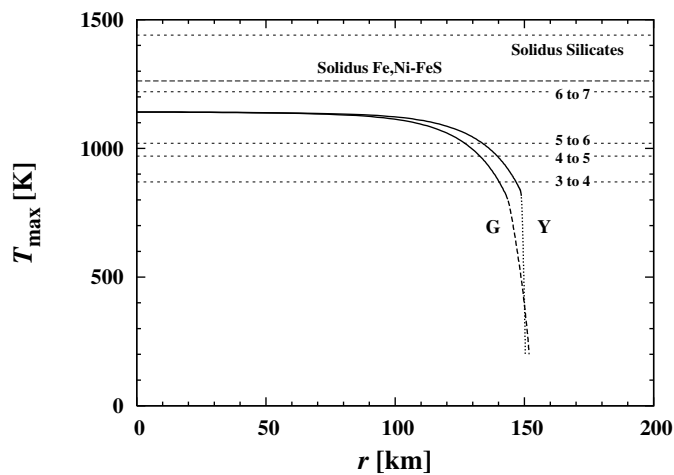}

\bigskip

\caption{Maximum value of the temperature at radius $r$ achieved during the whole thermal evolution of the body for models calculated with a porosity dependence of the heat conductivity for granular material (model G) as calculated in this work (long dashed) and for the dependence as derived from meteoritic data (model Y) of \citet{Yom83} (short dashed). The solid part of the lines correspond to that part of the body where the material is compacted by sintering. Also shown are the limits between the range of metamorphic temperatures for different petrologic classes and the solidus temperatures of metal and silicates.}

\label{FigTmax}

\end{figure}

For the model calculations the dependence of the average heat conductivity on porosity is calculated according to  Eq.~(\ref{IntKpor}). For the bulk heat conductivity $K_\mathrm{b}$ the value calculated for material with the average composition of H chondrites given in Table~\ref{TabKeffChond} is used. For the temperature dependence of $K_\mathrm{b}$ a provisional dependency $\propto T^{-1/2}$ is assumed, which is a reasonable approximation for rock material \citep[e.g.][]{Xu04}. A more detailed investigation of the temperature dependence will be published elsewhere. 

The heat production in the parent bodies is determined by the decay of short and long lived radioactives. The heat production by $^{26\!}$Al and $^{60}$Fe governs the initial period of the thermal evolution. Their initial abundances at time of CAI formation used in our calculations are given in Table~\ref{TabParmTest}. The reasons for this choice are discussed in \citet{Hen13}. The calculation also includes heating by long lived radioactives as described in \citet{Hen11}. They are important for the late thermal evolution.

\begin{table*}

\caption{%
Parameters for optimised models of the parent body of H chondrites using different approaches for calculating the heat conductivity of the porous material. Model G uses the heat conductivity for a granular chondritic material as derived in this paper. Model Y uses the porosity dependence of $K$ as found for impact fractured material. For comparison, model H shows the result from model H of \citet{Gai14b}. }

\begin{tabular}{ll@{\hspace{1.5cm}}l@{\hspace{0.8cm}}l@{\hspace{0.8cm}}ll}
\hline
\hline
\noalign{\smallskip}
         &       & Model G & Model Y & Model H &  \\
\cline{3-5}
\noalign{\smallskip}
Quantity & Symbol& Value & Value & Value & Unit \\
\noalign{\smallskip}
\hline
\noalign{\smallskip}
 & & \multicolumn{3}{c}{Parameters} & \\
\noalign{\smallskip}
Heat conductivity             & $K_{\rm b}$      & 4.9                & 4.9                & 4.0                & $\rm W (mK)^{-1}$\\[.2cm]
\noalign{\smallskip}
 & & \multicolumn{3}{c}{Optimised parent body} & \\
\noalign{\smallskip}
Radius                        & $R_{\rm p}$      & 154.7              & 168.9              & 159.4                & km \\
Formation time                & $t_{\rm form}$   & 1.940              & 1.879              & 1.840                & Ma \\
Surface temperature           & $T_{\rm srf}$    & 240.8              & 167.2              & 159.0                & K\\[.2cm]
Max. central temperature      & $T_{\rm c}$      & 1\,215             & 1\,240             & 1\,276                & K \\
Residual porous layer         &                  & 8.5                & 2.5                & 2.09                & km \\
\noalign{\smallskip}\hline
\noalign{\smallskip}
Fit quality &  \hspace{-.5cm}$\chi^2/(N-p)$      & 0.65               & 0.73               & 0.79                &  \\
\hline
\hline
\noalign{\smallskip}
Meteorite & type & \multicolumn{3}{c}{Burial depth and maximum temperature $T_{\rm max}$} & \\
\noalign{\smallskip}
          &      & km\hfill K & km\hfill K & km\hfill K &  \\
\noalign{\smallskip}
\hline
\noalign{\smallskip}
Estacado        & H6 & 60.0\quad  1\,191         & 46.6\quad 1\,202          & 40.4\quad 1\,235        &  \\
Guare\~na       & H6 & 57.6\quad  1\,190         & 43.9\quad 1\,201          & 38.8\quad 1\,234        &  \\
Kernouv\'e      & H6 & 42.2\quad  1\,180         & 29.9\quad 1\,186          & 27.1\quad 1\,217        &  \\
Mt. Browne      & H6 & 31.1\quad  1\,159         & 20.0\quad 1\,153          & 18.2\quad 1\,182        &  \\[.2cm]
Richardton      & H5 & 20.4\quad  1\,101         & 10.7\quad 1\,079          & 10.2\quad 1\,100        &  \\
Allegan         & H5 & 19.4\quad  1\,096         & 9.60\quad 1\,059          & 9.01\quad 1\,080        &  \\
Nadiabondi      & H5 & 16.4\quad  1\,057         & 5.46\quad\phantom{1\,}981 & 7.56\quad 1\,051        &  \\[.2cm]
Forest Vale     & H4 & 7.36\quad\phantom{1\,}851 & 1.53\quad\phantom{1\,}758 & 1.73\quad\phantom{1\,}766 &  \\
Ste. Marguerite & H4 & 5.67\quad\phantom{1\,}754 & 1.30\quad\phantom{1\,}674 & 1.69\quad\phantom{1\,}751 &  \\
\noalign{\smallskip}
\hline
\end{tabular}

\label{TabThermDat}
\end{table*}

\begin{figure*}

\includegraphics[width=\hsize]{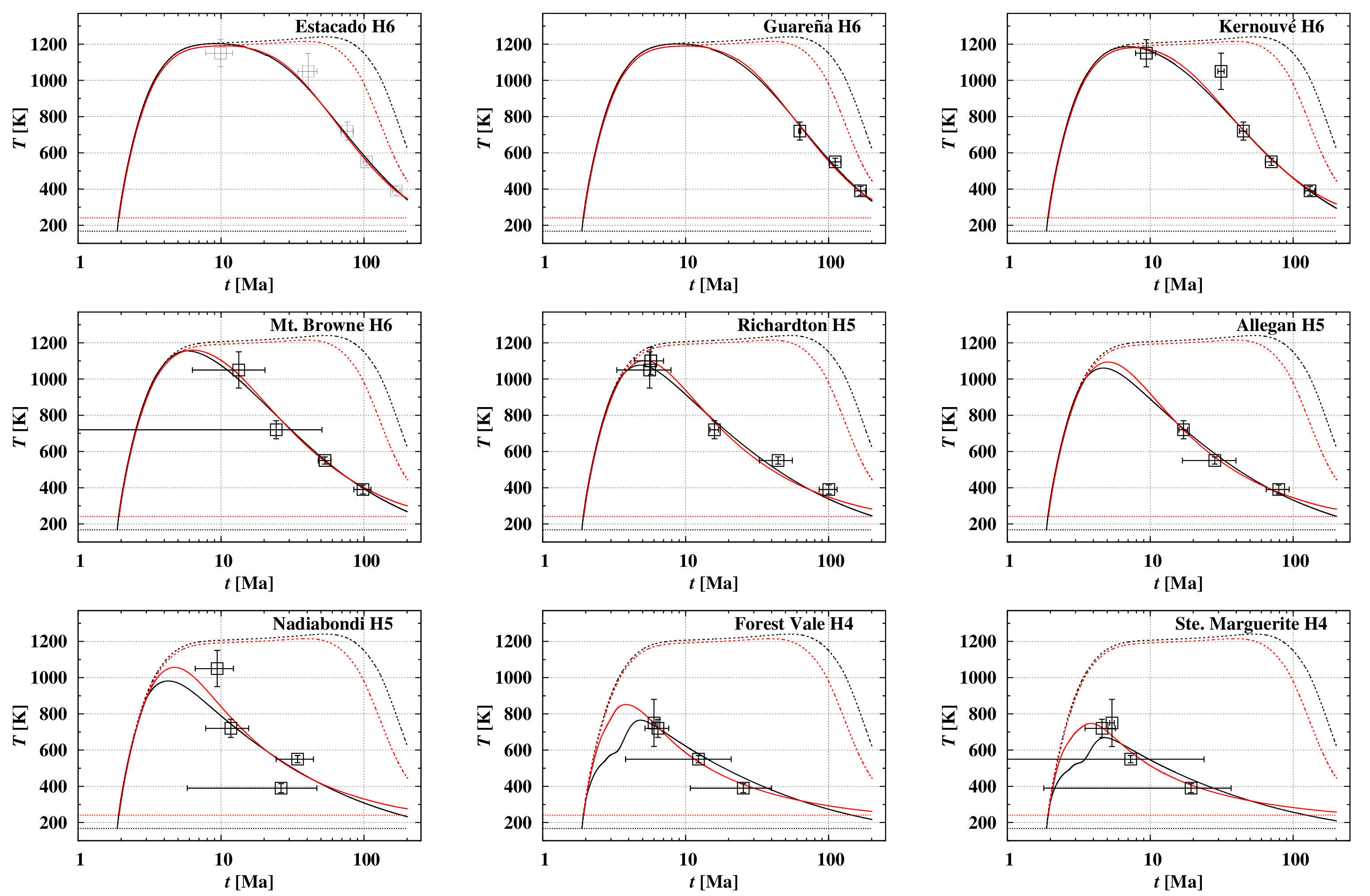}

\caption{Model fits to thermochronological data of the nine indicated meteorites of class H. Squares with errorbars indicate the empirical data for closures time of some radiometric clocks with different closure temperatures. Full red lines correspond to the temperature evolution at the derived burial depths of the individual meteorites for an optimized model of the parent body of the H chondrites   calculated with a heat conductivity of the chondritic material as calculated in this work. The full black line corresponds to the case that a prescription for the dependence of the heat conductivity on porosity  as proposed in \citet{Hen11} is used in the calculation. The short dashed lines show the time evolution of the central temperature of the body.  }
 
\label{FigEvoChondH}

\end{figure*}

\subsubsection{Sample calculation}

We compare models for the thermal evolution of a pla\-netesimal of radius 150\,km and formation time 2 Ma after CAI calculated with two different assumptions with respect to the porosity dependence of the heat conduction. The surface temperature is set in all cases to 200\,K. We denote the models using the porosity dependence of heat conductivity as for granular material as model G and with porosity dependence as derived from the meteoritic data of \citet{Yom83} and \citet{Ope12} as model Y.

Figure \ref{FigTestMod} shows the temperature evolution at depths of 3, 10, and 30 km beneath the surface and the temperature evolution at the centre of the body. The temperature evolution in both cases is significantly different due to the strongly different blocking of heat flux by pores in a granular material and in cracked material. Though the heat conductivity is identical after all voids are closed by sintering, the structure of the outer layer where temperatures never became sufficient for sinter-compaction of the material is different for both cases. The surviving porous layer extends much deeper in the case where we apply the porosity dependence of $K$ for granular material than for the case of cracked meteoritic material. Correspondingly, the thermal shielding effect of the outer porous layer is less marked for the granular layer because of its higher heat conductivity. This affects the temperature structure in the model in the sense that, except for the shallowest layers, 
 the descending part of the temperature curve at a certain depth is shifted in the model for the granular case by about 30 Ma to earlier times compared to the case of impact fractured material. 

Figure \ref{FigTmax} shows the resulting maximum temperature achieved at some radius during the thermal evolution of the body. This determines the compaction and thermal metamorphism of the material as it is reflected by the different petrologic classes of meteorites. One recognizes the significant differences in the depth of the surviving porous outer region between the two kinds of models. The larger extent in model G of the porous outer layer from which meteorites of petrologic type 3 originate appears to be more realistic than the very thin layer in model Y, because the outermost 1 km surface layer of a small body are probably significantly eroded by impacts during the 4.5 Ga evolution of the asteroid belt as may be inferred from the models in \citet{Hou79}. 

One conclusion we can draw from Fig.~\ref{FigTestMod} is that the temperature evolution inside the body as far as it is of interest for the thermo-chronometers is completed within about 100 Ma. If during this period a disruptive collision occurs, this completely changes the temperature history of the remaining fragments, even if they re-assemble to a rubble pile. If no such catastrophic event occurs within the first 100 Ma, strong impacts may result in strong local disturbances of the thermal evolution of the matter, but they do not result in a global change of the thermal history of the material. If disruptive impacts occur after the critical initial evolutionary phase of about 100 Ma, the bigger fragments should preserve in their interior the information on the initial  temperature evolution. Then at least some fraction of the meteorites derived from the bigger fragments preserves the record on the thermal history within the crucial initial $\sim100$ Ma period.   

For a crude estimation of the effect that a growing im\-pact-modi\-fied outer layer has for the thermal evolution we consider a very simple approximation. We follow the results of \citet{Hou79} for a body with 300 km diameter and strength scaling and assume that a regolith layer grows linear in time to 2 km thickness within 200 Ma. An initial porosity in this layer is assumed to equal 0.25 as for the granular material. The porosity dependence of the heat conductivity in this impact generated covering layer is assumed as for meteoritic matter (model Y). Figure \ref{FigTestMod} shows the temperature evolution for the centre and at three selected depths in such a model. The temperatures in this model are close to that in model G for the initial $\sim 100$\,Ma year of evolution and only for later times start to deviate somewhat. In all cases the deviations remain $\lesssim30$\,K for temperatures $>400$\,K which is less than the accuracy with which the closure temperatures of thermochrono
 meters are known (see Fig.~\ref{FigEvoChondH} or Table 5 of \citet{Gai14b} for closure temperatures of the thermo-chronometers used in this work). The closure times derived from thermochronometers therefore seem not to be significantly influenced by a gradually growing impact-generated regolith layer. For the moment we feel justified by this finding to ignore the whole subject, but the impact evolution of the surface layers clearly needs a detailed study, which is out of the scope of this paper. 
 
\subsection{Fit of thermochronological data of H chondrites}

We apply our calculation of the heat conductivity of chondritic material to the problem of reconstructing the parent body of the H chondrites from thermochronological data of a number of selected chondrites of class H. A model is searched for that reproduces the thermal history of a number of H chondrites for which their individual cooling history is sufficiently constrained by empirical data. We have discussed this problem in our earlier papers \citep{Hen11, Hen12, Hen13, Gai14b} and will not repeat the details. We base our calculation on the same set of 9 meteorites and 37 data for closure ages as described in \citet{Gai14b}. There are still only nine meteorites available which satisfy the requirement that their individual cooling history is determined by closure ages for at least three different thermo-chronometers.

The basic pysical parameters ($G$, $\phi_0$, etc.) assumed for the models are given in Table~\ref{TabParmTest}. The basic parameters determining the properties of an evolution model of a planetesimal (radius $R_\mathrm{p}$, formation time $t_\mathrm{form}$, surface temperature $T_\mathrm{srf}$) are determined by comparison with empirical cooling histories of meteorites. For a given model the unknown burial depth within the parent body from which the meteorites originate are determined for each meteorite by an optimisation method such that the temperature evolution at the burial depth fits as close as possible to the thermochronological data points of that meteorite. The parameters determining the evolution model of the parent body are varied by applying an evolution algorithm such that an overall best fit for the set of all meteorites is obtained in the sense that the calculated deviation $\chi^2$ between model and data is lowest. 

The result of the optimisation process is shown in Table~\ref{TabThermDat} and in  Fig.~\ref{FigEvoChondH} as model G. For comparison we ran the same optimisation with the porosity dependence of the heat conductivity as determined for meteorites (model Y). The first group of data in the table show the optimized values of the model parameters ($R_\mathrm{p}$, $t_\mathrm{form}$, $T_\mathrm{srf}$) and the resulting maximum central temperature and the thickness of the residual porous layer. For comparison also the best-fit model H from \citet{Gai14b} is shown which differs from model Y only in so far, as not the calculated bulk heat conductivity $K_\mathrm{b}$ of the material of H chondrites (Table \ref{TabKeffChond}) but some estimated value is used.

The radius and formation time of the models are quite similar for all the models. They are essentially constrained by the cooling history at large depth, i.e., by the data for the meteorites of high petrologic type. For the deep layers of the body the temperature evolution is almost the same, as is readily seen from Fig.~\ref{FigEvoChondH}. The differences are more pronounced at shallow outer layers where the structure of the model is different for model G and Y. As already shown by the sample calculations, the residual porous layer in model G is significantly thicker than in Model Y as result of the higher value of the heat conduction coefficient for sintered granular material than for a material where heat flux is efficiently blocked by cracks. The temperature lapse rate close to the surface therefore is shallower in model G than in model Y and the temperature for efficient sintering of the chondrules at $\gtrsim 900$\,K is encountered only at higher depth in model G than in model 
 Y. 

The strong difference in heat conductivity in the porous outer layer is also the reason for the higher surface temperature required in model G than in model Y because otherwise cooling would be too rapid in the layers where the H4 chondrites come from. The cooling history of the low petrologic classes mainly constrains the surface temperature.    

On the whole, the model using the new model for heat conductivity of pristine chondritic material results in a somewhat better fit quality, expressed by a lower value of $\chi^2/(N-p)$ ($N$ = number of data, $p$ = number of parameters). This supports to some extent our hypothesis, that the heat conductivities measured for meteoritic material cannot be applied to the early evolutionary phase of the parent bodies, because the micro structure of the material is strongly impact modified over the long period between completion of the initial heating- and cooling phase and the investigation of the meteorite in the laboratory.

\section{Concluding remarks}
\label{SectConclu}

We have studied in this work the heat conductivity of chondritic material. It is shown that it is possible to determine for ordinary chondrites the effective heat conductivity coefficient from the individual properties of the components of the complex mixture of minerals and metal in chondrites by solving the heat conduction equation for a test body. We found that the effective heat conductivity of the compact material can be approximated with sufficient accuracy by the Bruggeman mixing rule. This reproduces the result of the numerical calculation with an accuracy of better than one percent. Hence the value of the heat conductivity coefficient of compact material can be calculated equally well from the Bruggeman mixing rule.

We have also studied the effective heat conductivity of porous material by numerically solving the heat conduction equation for a test body. We considered a model with a stochastic distribution of voids and a model where a granular material formed from spheres with some different mixtures of sphere radii is numerically generated and artificially sintered. It is found that the variation of effective heat conductivity strongly deviates from what is found by laboratory measurements of heat conductivity of meteorites, while the calculations can reproduce measured values for natural sandstone. It is argued that the discrepancy can be explained by the fact that the pore space in meteorites is dominated by impact generated cracks which have a sheet-like structure differing completely from the geometry of voids in compacted granular  material formed from spheres. It is demonstrated by application of Bruggeman mixing theory for disk-like voids in a porous material that the difference between 
 calculated and measured heat conductivities can be readily explained by micro-cracks in meteoritic material.

It is argued then that meteoritic material does not represent the pristine material present in the earliest evolutionary phase of the parent bodies of asteroids, because  according to theory the build-up of a heavily impact modified surface layer where numerous cracks are generated in the material requires timescales longer than the critical period for the thermal evolution of the parent bodies of chondrites. Laboratory data for heat conductivity of chondritic material are only representative for present day asteroids. The pristine material present during the initial thermal evolution seems no more to exist and one has to determine its properties from calculations, as we did, or one could produce in the laboratory a crack-free material and measure its properties.

Our calculations show that also the porosity dependence of the granular material may be calculated with no more than about ten percent deviation from the numerical results from the Bruggeman mixing rule. On the other hand, the behaviour of the meteoritic material may be modelled by the Bruggeman mixing rule for disk-like inclusions. We also give an analytic approximation to our numerical results which may be applied in model calculations.

We repeated our earlier fits of thermal evolution models for the H chondrite parent body with thermochronological data for the set of nine H chondrites with rather well defined cooling histories and determined the putative radius and formation time of the body. The results are close to our earlier results based on an effective heat conductivity as measured for meteorites. With the new data on heat conductivity the quality of the fit is, however, somewhat improved. 
    
The discrepancy between the results of laboratory measurements and theoretical predictions for the effective heat conductivity strongly suggests that attention has to be paid to the question what the properties of the pristine material in planetesimals are. It is obviously necessary to study in detail the impact evolution of the surface layers before one can come to a final conclusion on the thermal history of meteorites and their parent bodies. 

The model for porosity of meteoritic material used in this paper cannot be applied to the uppermost few meters below the surface where space weathering is active. There one probably encounters material properties which are not made allowance for in our model \citep[e.g., a very high porosity, ][]{Eme06,Ver12} and that requires a different treatment.
   

\begin{acknowledgements}
We greatly acknowledge that E. Beitz, F. M\"ohlmann, and J. Blum shared their data on the structure of the sintered body of glass-beads with us. This work was performed as part of a project of `Schwerpunktprogramm 1385', supported by the `Deutsche Forschungs\-gemeinschaft (DFG)'. This research has made use of NASA's Astrophysics Data System.
\end{acknowledgements}


\begin{appendix}

\section{The numerical model}
\label{NumModHeat}

The heat conduction equation is numerically solved for a cube of edge-length $L$. We chose for this purpose an implicit finite differences method which is about an order of thousand times faster on a single cpu than an explicite method since there is no restriction on the time step length in implicit methods to guarantee stability. The cube is divided into a grid of $n^3$ equal sized sub-cubes of edge length $h=L/n$, with $n$ being the number of grid cells in each dimension.

Writing equation (\ref{WLG}) in implicit finite differences yields
\begin{equation}
T^{n+1}_{i\!jk} - \left( \frac{1}{\rho c_{\!p}} \right)_{i\!jk} \cdot \left( \nabla K \cdot \nabla T \right) _{i\!jk}^{n+1} \cdot \Delta t = T^{n}_{i\!jk} ,
\end{equation}
with $\Delta t$ being the time step size and
\begin{align}
\left( \nabla k \cdot \nabla T \right)_{i\! jk} =& \\[.2cm]
&\hskip-1.3cm 
\left ( \frac{ K_{i+1/2} T_{i+1} + \left( K_{i-1/2} - K_{i+1/2} \right) T_{i} + K_{i-1/2} 			T_{i-1} }{ h^2 } \right)_{\!jk} \\[.2cm]
&\hskip-1.6cm
+\left ( \frac{ K_{\!j+1/2} T_{\!j+1} + \left( K_{\!j-1/2} - K_{\!j+1/2} \right) T_{\!j} + K_{\!j-1/2} 			T_{\!j-1} }{ h^2 } \right)_{ik} \\[.2cm]
&\hskip-1.6cm
+\left ( \frac{ K_{k+1/2} T_{k+1} + \left( K_{k-1/2} - K_{k+1/2} \right) T_{k} + K_{k-1/2} 			T_{k-1} }{ h^2 } \right)_{i\!j}.
\label{Differenzengleichung}
\end{align}
Here the index $n$ denotes the $n$th time step of length $\Delta t$, and $i,j,k$ the indices of the spatial coordinates in each spatial direction, and $ h = L/n $ denotes the spatial step size (grid constant). Equation (\ref{Differenzengleichung}) results in a large and sparse coupled system of equations with $n^3$ unknowns which is solved by a conjugate gradient method. 

The quantities $K_{i \pm 1/2}$ denote the heat conductivity at the midpoint of the interval between index $i$ and that of index $i\pm 1$. One possibility is to define this as the arithmetic mean
\begin{equation}
K_{i\pm 1/2}= (K_{i\pm 1}+K_{i})/2\,,
\end{equation}
another possibility is the harmonic mean
\begin{equation}
K_{i\pm 1/2}= 2 / (K_{i\pm 1}^{-1}+K_{i}^{-1})\,.
\end{equation}
The harmonic mean is a more appropriate treatment if the heat conductivities of the different components used vary significantly but cannot be used if one component has a heat conductivity of zero. In this paper, the harmonic mean was used in all cases where the heat conductivity of chondritic material is calculated. For the calculations of the heat conductivity of systems of spheres in a vacuum, the arithmetic mean is used.

\begin{figure}

\includegraphics[width=\hsize]{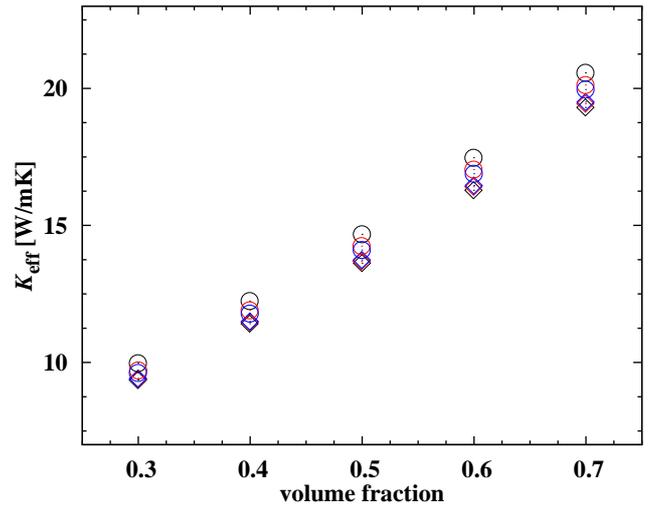}

\caption{Comparison of effective heat conductivity $K_{\rm eff}$ for different volume fractions of iron in an  olivine matrix using the arithmetic and harmonic mean. Circles: arithmetic mean with grid resolution $n$=100, 200, and 300  from top to bottom. Diamonds: harmonic mean with $n$=100, 200, and 300 from bottom to top.}

\label{FigCompKipDiscr}

\end{figure}

Figure~\ref{FigCompKipDiscr} shows a comparison of the resulting effective heat conductivites resulting from the two means for a two component material with different component mixing ratios. Plotted are results of both means at grid resolutions of   $n=100$, 200, and 300. It is easily seen that the results for both means approach each other for increasing resolution. And this approaching is mainly by the solution based on the arithmetic mean, whereas the $K_{\rm eff}$ for the harmonic one stays basically the same. From this behaviour we conclude that using the harmonic mean in our models is the better choice if we do not consider areas with zero heat conductivity.


As boundary conditions for solving the equation we have constant temperatures $T_1$ and $T_2$ with $T_1 \neq T_2$ on two opposite sides of the cube, i.e., a Dirichlet boundary condition. At the other four sides we describe Neumann boundary conditions. This boundary condition assumes that there is no effective heat flux through the side walls of the cube. This is certainly not correct for an individual grain located at the boundary because of certain local fluctuations of the heat flux. Strictly speaking this boundary condition can only apply on the average over larger pieces of the boundary area, i.e., after we have already performed the averaging procedure. In practice, however, we wish to consider a situation where averaged quantities vary over length scales which  are much larger (kilometers) than typical sizes of individual particles ($\mu$m to mm). In that case local deviations of heat flux from the mean must be very small such that a Neumann boundary condition appears to be a r
 easonable approximation.

As initial condition any temperature distribution throughout the cube can be prescribed that fulfills the boundary conditions. Here we prescribe the mean between $T_1$ and $T_2$ for each grid cell not belonging to the front surfaces with constant boundary temperature.

To determine the effective heat conductivity this model is then run until a stationary state establishes for the temperature distribution. We assume a stationary state to be achieved, if the maximum temperature change in any grid cell for a time step is less then $10^{-7}$ K/s.
In that stationary state the heat flow will not change any more and can be used to determine the effective heat conductivity $k_{\rm eff}$ Eq. (\ref{DefEffK}). The total heat flow $F$ through the front sides can be determined by summing over the heat flows of each grid cell that belongs to a certain boundary surface of constant temperature.

The model can be run for any resolution, whereas a resolution of $200 \times 200 \times 200$ is a good compromise between a high enough resolution and a reasonable computation time. For a cube with $L$ = 1 cm and an effective heat conductivity of the order of 10 W/mK the physical time until a stationary state establishes is of the order of 1 min.


\section{Random distribution of materials}
\label{AppRanDi}

A random distribution of components with given volume fractions is generated via the following procedure:

1. The material is supposed to consist of $N$ components. These components are to be distributed randomly within the cube with a prescribed volume fraction for each component.

2. In the first step, the cube is decomposed into sub-cubes. The number of grid cells, $n$, for each direction is chosen and the set of volume elements of the decomposition is generated. All grid-cells are supposed to be filled with one of the components, now called component~1. 

3. In the second step, centers of equal sized balls with a prescribed radius $R$ are generated at random by using a random number generator. For all grid cells for which their centre lies within such a ball its component is changed to component 2. The balls are generated one by one until the number of cells filled with component 2 attains the prescribed volume fraction of this component.

This will result in a cube with a matrix component (component 1) containing some balls consisting of the second component. In this procedure overlapping of balls is not excluded, resulting in arbitrary shapes of component 2 aggregates within the cube if the number of balls and their radius is such that there is significant overlap.

4. If there are three or more components, the preceding step is repeated for each of the further components. For each grid cell where its centre is located in one of these new balls and its component is still component 1, its component is changed to the new component.

This procedure allows to generate a random spatial distribution of a number of different materials inside the cube with a prescribed volume fraction of each component. By means of construction, the prescribed volume fraction can only approximately be fitted because of the final increments of component volume during the construction process. Different constructions starting with different seeds of the random number generator give slightly different volume fractions around its nominal value. This resembles what one observes in the compacted material of chondrites of petrologic types 5 and 6 and we take such artificially generated distributions as a model for compacted chondritic material.
 
The distribution of components generated in this way does not depend on the grid resolution of the finite difference method but only on the assumed radius $R$ of the balls and the volume fractions of the components. Thus, starting with the same sets of balls for the different components, the same spatial distribution of components with different decompositions of the cube into $n^3$ grid cells can be generated for calculating the effective heat conductivity by varying $n$. In addition, the radius $R$ of those balls can be varied to define component structures of varying sizes.


\section{%
Construction of a granular medium}
\label{NumModGran}

A box filled with balls either of a fixed radius $R$ or with radii from some assumed radius distribution is considered.  

A rectangular Cartesian coordinate $x$, $y$, $z$ is used. The coordinates of the walls perpendicular to the $x$, $y$ and $z$ axis are $X_0$, $X_1$ and $Y_0$, $Y_1$ and $Z_0$, $Z_1$, respectively. The width $D$ of the box in the $x$ and $y$ direction have to be chosen as a suited multiple of the (average) ball radius. In case of fixed ball radii, $D$ must not be a multiple of $R$ because the system tends to form an unwarranted cubic sphere packing in this case. The height of the box in $z$ direction must be at least twice the width $D$.

It is assumed that there is some downward acceleration
\begin{equation}
A_z^{(\rm g)}=-g
\end{equation}
acting on the balls and that there is some frictional acceleration
\begin{equation}
A_x^{\rm(fr)}=-u_x/\tau\,,
\end{equation}
where $u_x$ is the $x$ component of the velocity and $\tau$ a given stopping time, and analogously for the other two directions. The balls experience a repulsive acceleration on contact given by
\begin{equation}
A_{x,ij}^{\rm(r)}=\begin{cases}\alpha_{\rm b}{x_i-x_j\over\rho_{ij}}&\mbox{if}\quad \rho_{ij}<R_i+R_j\\
0 &\mbox{else}
\end{cases}
\end{equation}
and analogously for the other two components, where
\begin{equation}
\rho_{ij}=\left((x_i-x_j)^2+(y_i-y_j)^2+(z_i-z_j)^2\right)^{1/2}
\end{equation}
and $x_i$, $y_i$, \dots\ are the coordinates of the centres of the balls, $R_i$ and $R_j$ the radii of the balls, and $\alpha_{\rm b}$ is a suited constant. If the distance of a ball centre to one of the walls is less than the radius, the same kind of repulsive acceleration
\begin{equation}
A_{x,i}^{\rm(w)}=\alpha_{\rm w}{x_i-x_j\over R_i}
\end{equation} 
is applied.

The following equations are solved for the balls
\begin{equation}
\ddot x=A_z^{(\rm g)}+A_x^{\rm(fr)}+A_{x,ij}^{\rm(r)}A_{x,i}^{\rm(w)}
\end{equation}
with a simple leap-frog method. The balls are introduced one after another at the ceiling of the box with random $x$, $y$ coordinates at short intervals. 

One of the sidewalls and the bottom are vibrating, e.g., the $X_0$ and $Z_0$ coordinate are periodically moved with slight amplitude. This simulates a shaking of the box which in turn induces a chaotic motion of the balls. The shaking is gradually switched off some time after the last ball is introduced. At the same time the position of the ceiling is gradually lowered until it fits to the upper layer of balls.

\begin{table}

\caption{%
Parameters for generating random packing of balls.}

\begin{tabular}{cccccc}
\hline
\hline
\noalign{\smallskip}
$R$ & $D$ & $g$ & $\tau$ & $\alpha_{\rm b}$ & $\alpha_{\rm w}$\\
\noalign{\smallskip}
\hline
\noalign{\smallskip}
0.52 & 5 & 0.1 & 10 & 100 & 1000 \\
\noalign{\smallskip}
\hline
\end{tabular}

\label{TabAppRB}
\end{table}

The system evolves by friction to a final state where all balls are at rest. Rather big time steps are taken for integrating the equations of motion, because it is not necessary to achieve an accurate orbit of the balls; the assumed forces are anyhow artificial. Only the final stage is of interest after all balls have come to a rest. This procedure results in a random packing of balls. The parameter values typically used in a run are given in Table~\ref{TabAppRB}.

\end{appendix}


\end{document}